\documentstyle[aps,prl,multicol]{revtex}

\include{epsf}
\begin{document}
\preprint{\today}
\draft 
\title{\bf Molecular Correlations 
in a Supercooled Liquid}

\author{L. Fabbian$^{(1)}$, A. Latz$^{(2)}$, R. Schilling$^{(2)}$, 
F. Sciortino$^{(1)}$,  P. Tartaglia$^{(1)}$, C. Theis$^{(2)}$}

\address{$^{(1)}$ Dipartimento di Fisica and Istituto Nazionale
 per la Fisica della Materia, Universit\'a di Roma {\it La Sapienza},\\
 P.le Aldo Moro 2, I-00185, Roma, Italy \\
$^{(2)}$ Institut f\"ur Physik, Johannes Gutenberg--Universit\"at,
Staudinger Weg 7, D-55099 Mainz, Germany }

\date{\today} \maketitle

\begin{abstract}

We present static and dynamic properties
of molecular correlation functions $S_{lmn,l'm'n'}(\vec{q},t)$ in 
a simulated supercooled liquid
of water molecules, as a preliminary effort in the direction of
solving the molecular mode coupling theory (MMCT) equations for 
supercooled {\it molecular} 
liquids. The temperature and time dependence of various 
molecular correlation functions,  calculated from 250 $ns$ long 
molecular dynamics simulations, show the characteristic patterns 
predicted by MMCT and shed light on the driving mechanism 
responsible for the
slowing down of the molecular dynamics.
We also discuss the symmetry properties of the  
molecular correlation functions
which can be predicted on the basis of the
$C_{2v}$--symmetry of the molecule. The analysis of the 
MD--results for the static correlators $S_{lmn,l'm'n'}(\vec{q})$
reveals that additional relationships between
correlators with different signs of $n$ and $n'$ exist. 
We prove that for molecules with
$C_{rv}$--symmetry  this unexpected result becomes exact at least for
high temperatures.

\end{abstract}


\begin{multicols}{2}

\section{Introduction}
\label{introduction}

In recent years, significant progress has been made in the
understanding of the slow dynamics in supercooled liquids
\cite{confproc}.
Theoretical\cite{gotze,schilling94,parisi,coniglio}, experimental
(see e.g. \cite{hsexp,cummins,tolle,richter,yip}) 
and simulation efforts (see e.g. \cite{signorini,barrat,kob,lw}) have 
highlighted 
the role played by the ideal-glass transition temperature $T_c$, first
predicted by mode-coupling theory (MCT) for {\em simple} liquids 
\cite{BGS,mazenko} and in a schematic model \cite{leutheusser}, and 
identified the different dynamical mechanisms
above and below this temperature.  This work has clarified the strong
interplay between liquid structure and liquid dynamics above $T_c$ as
well as the universal aspect of the decorrelation process --- which
close to $T_c$ is predicted to become independent of the correlation
function as well as the $q$ vector that is probed\cite{leshouches}.
For colloidal systems MCT has been tested both on a
semi--quantitative level \cite{hsexp}, 
i.e. the validity of the scaling laws was
investigated, and also on a quantitative level where it has been
demonstrated that the time--dependent density correlator for a liquid
of hard spheres (which is a good model for neutral colloids) obtained
from MCT describes the corresponding experimental result over 3
decades in time by using {\em only one} fit parameter (for the
timescale) \cite{megen,arnulf}. All the other experimental tests 
were semi--quantitative
and were exclusively restricted to {\em molecular} systems, like OTP,
Salol, Glycerol, etc. (for more details see refs. 
\cite{gotze,schilling94}, \cite{cummins,tolle,richter,yip,note} and
references therein).  On a first glance the reasonably
good agreement between the predictions of MCT for {\em simple} liquids
with many of the experimental data for the molecular glass forming
liquids seems to be surprising, since the orientational degrees of
freedom do not appear in the original version of MCT 
\cite{leshouches}.
%Theoretical predictions for the time dependence of orientational degrees of freedom in supercooled
%liquids demands an extension of MCT to molecular
%liquids. 
To do a quantitative comparison between theory and experiment or 
simulation and especially to describe the molecular correlations 
(including orientational degrees of freedom) it is necessary to extend MCT to molecular liquids.
In particular, a molecular mode
coupling theory (MMCT) allows to study the role of the coupling
between the translational degrees of freedom and orientational degrees of freedom. Such an
extension has recently been performed for a single dumbbell molecule
in an isotropic liquid \cite{dumbbell} 
and for a liquid of diatomic molecules
\cite{schilling,theisdipl,kawasaki}. Reference \cite{theisdipl} and 
\cite{kawasaki} even treat general molecules.

MMCT is also
conceptually based on the hypothesis that the liquid structure is
controlling the long-time dynamical evolution of the system.
Structural information is used as input in the theory via
generalization of the density-density structure factor including the
angular degrees of freedom.  Such quantities, although they are 
difficult to determine experimentally, can be evaluated from molecular
dynamics trajectories and used to test the quality of the recently
proposed MMCT. For the case of a liquid of diatomic 
molecules, the molecular structure factors have 
been evaluated\cite{kks} and
a first quantitative test of MMCT has been presented in Ref.
\cite{theis98,winkler}.
Furthermore, the temperature dependence of the molecular 
correlation functions 
can give hints regarding the leading mechanism for the slowing
down of the dynamics.

The asymptotic predictions of MCT continue to be valid within the
MMCT scheme. Also in MMCT\cite{dumbbell,schilling} 
$i)$ there exists a $\beta$--relaxation regime where the factorization
of the time-- or frequency--dependence of the correlators or
susceptibilities from the space-- and angular dependence holds
generically and the time--dependence is given by the
$\beta$--correlator which fulfills the {\em first scaling law}
\cite{leshouches} and $ii)$ there exists an 
$\alpha$--relaxation regime in
\end{multicols}
\twocolumn
\noindent which
the {\em second scaling law} \cite{leshouches} holds, i.e. 
the 
time--temperature
superposition principle is fulfilled\cite{predictions}. 
These two results 
underline the {\em universality} of the ideal structural glass
transition based on a bifurcation scenario described by a
fold--singularity \cite{leshouches}. 
However, the calculation of the various exponents (entering the
scaling laws), the calculation 
of the critical
nonergodicity parameter, the critical amplitudes, the transition
temperature $T_c$ and particularly of the molecular correlation
functions themselves require the solution of the  MMCT equations.

In this Article we present the temperature and $q$-vector dependence
of the generalized structure factors required by the MMCT for the case
of SPC/E potential\cite{spce}. 
This potential, which describes the molecule as a
rigid planar body and models the pair interactions as a sum of
electrostatic and Lennard Jones terms, has been studied in detail and
has been shown to reproduce qualitatively the characteristic
properties of liquid water---a liquid where the slowing down of the
dynamics on cooling
 is {\em not} related to packing constraints but to
the formation of a tetrahedral network of highly directional hydrogen
bonds\cite{lavorisuspce,Rocca}.
Previous numerical 
studies on the SPC/E system, focused on the center of mass  
self\cite{self} and collective properties\cite{collective}, 
have shown on a semi--quantitative level that the center of mass dynamics 
is well described by MCT, with an estimated critical temperature of
about $T_c=200 \pm 3 K$. It has been shown that
orientational degrees of freedom are crucial in strongly enhancing the slowing
down of the dynamical processes on supercooling. 
This has led to the introduction of a semi--schematic 
MCT model \cite{semischem} where the coupling to the
orientational degrees of freedom is accounted for phenomenologically by the introduction of a
parameter $\chi_R \ge 1$ by which the coupling of the center of mass--density
modes is enhanced. 
This semischematic approach, recently tested also for the case of 
a supercooled liquid of diatomic molecules\cite{schem-dumb},
focuses only on the center of mass translational degrees of freedom and is
by construction unable to describe the time evolution of
the orientational degrees of freedom and to indicate which orientational degrees of freedom are responsible of the 
slowing down of the center of mass and orientational dynamics. 
A MMCT description is required
to fully describe these important dynamical aspects.

The  generalized structure factors presented in this Article are 
a first step in the direction of solving the MMCT for a molecular 
system. In section \ref{sec:mcf} we motivate the introduction 
of an infinite--dimensional correlation
matrix $S_{lmn,l'm'n'}(\vec{q},t)$ where $l = 0,1,2,... \; ; \; -l \le
m \le l \; , \; -l \le n \le l$ and corresponding relations 
for the primed quantities.  As
already mentioned above,  MMCT requires the static correlators
$S_{lmn,l'm'n'}(\vec{q})$ as an input.  
In Sec.\ref{secsymmetry} we discuss
the relations between the $S_{lmn,l'm'n'}(\vec{q},t)$ 
correlation functions arising from 
the symmetries characteristic of isotropic liquids and the symmetries
characteristic of $C_{rv}$ molecules. 
These relations reduce drastically the
number of independent $S_{lmn,l'm'n'}(\vec{q},t)$ correlation functions
and support the feasibility of a full MMCT calculation for 
this class of liquids.
Finally, in Sec.\ref{sec:res} 
we present the time evolution of the generalized angular  
correlation functions  for the SPC/E case 
and interpret their behavior in the general MCT framework.  The
qualitative and semi-quantitative agreement between numerical data and
asymptotic MCT 
theoretical predictions strongly suggests to perform a full MMCT
comparison.

\section{Molecular Correlation Functions}
\label{sec:mcf}
To describe a molecular liquid it is necessary to introduce,
besides the information on the position of the 
molecule's center of mass,
the information on the orientation of the molecules. 
The microscopic density,  defined for
simple (atomic) liquids\cite{H&McD} as $\rho(\vec{x},t)= \sum_{j=1}^N 
\delta(\vec{x}-\vec{x}_j(t)) $, is  generalized to 
\begin{equation}
\rho(\vec{x},\Omega,t)= \sum_{j=1}^N \delta(\vec{x}-\vec{x}_j(t)) 
\delta(\Omega-\Omega_j(t))
\end{equation}
where the sum runs over the $N$ molecules of the liquid, 
$\vec{x}_j(t)$ and $\Omega_j(t)=(\phi_j(t),\theta_j(t),\chi_j(t))$ are
respectively the position of the center of mass and the Euler angles of the 
$j$th molecule at time $t$.

Any function $f(\vec{x},\Omega)$ can be expanded with respect to
plane--waves and to generalized spherical harmonics
$D^l_{mn}(\Omega) = e^{-im\phi} \;
d^l_{mn}(\theta) \; e^{-in\chi}$ (see Ref. \cite{G&G}) 
as 
\begin{eqnarray}
f(\vec{x},\Omega) &=& \frac{1}{8\pi^2V} \sum_q \sum_{lmn} (-i)^l 
(2l+1)^\frac{1}{2} \cdot \nonumber \\ 
& & \cdot \; \; f_{lmn}(\vec{q}) \; e^{-i \vec{q} \vec{x}} \;
D^l_{mn}(\Omega)
\label{a1}
\end{eqnarray}
where the coefficients $f_{lmn}(\vec{q})$ are given by
\begin{equation}
\label{a2}
f_{lmn}(\vec{q}) = i^l (2l+1)^\frac{1}{2} \int d^3x \int d\Omega
f(\vec{x},\Omega) \; e^{i \vec{q} \vec{x}} \; D^{l \ast}_{mn}(\Omega)
\end{equation}
Here it is $l \ge 0, \; -l \le m \le l, \; -l \le n \le l$.
Application of (\ref{a2}) to $\rho(\vec{x},\Omega,t)$ yields the
tensorial one--particle density:
\begin{equation}
\label{a3}
\rho_{lmn}(\vec{q},t) = i^l (2l+1)^\frac{1}{2} \sum_{j=1}^N \; e^{i
\vec{q} \vec{x}_j(t)} \; D^{l \ast}_{mn}(\Omega_j(t)).
\end{equation}
The prefactor in eq.(\ref{a3}) is chosen for technical convenience,
e.g. the factor $i^l$ makes the molecular correlators real for $n=n'$. 
As will be shown below, this choice  produces real correlators also for
$n$ and $n' \neq 0$ at high temperature. 
Now we can introduce the time-dependent {\em
molecular} correlation functions
\begin{equation}
\label{eq:sq}
S_{lmn,l'm'n'}(\vec{q},t) = \frac{1}{N} \langle
\rho_{lmn}^{\ast}(\vec{q},t) \; \rho_{l'm'n'}(\vec{q},0) \rangle
\end{equation}
where the angular brackets denote the canonical average over the
initial point in phase space. Substitution of (\ref{a3}) into
(\ref{eq:sq}) yields
\begin{equation}
\label{a5}
S_{lmn,l'm'n'}(\vec{q},t) = S_{lmn,l'm'n'}^{(d)}(\vec{q},t) +
S_{lmn,l'm'n'}^{(s)}(\vec{q},t)
\end{equation}
with the {\em distinct part}
\begin{eqnarray}
& & S_{lmn,l'm'n'}^{(d)}(\vec{q},t) = i^{l'-l} \left[ (2l+1)(2l'+1)
\right]^\frac{1}{2} \frac{1}{N} \sum_{j \not= j'} \cdot \nonumber \\ 
& & \;\; \cdot \quad \langle e^{-i
\vec{q} ( \vec{x}_j(t) - \vec{x}_{j'} )} \; D^l_{mn}(\Omega_j(t)) \;
D^{l' \ast}_{m'n'}(\Omega_{j'}) \rangle 
\label{a6}
\end{eqnarray}
and the {\em self part}
\begin{eqnarray}
& & S_{lmn,l'm'n'}^{(s)}(\vec{q},t) = i^{l'-l} \left[
(2l+1)(2l'+1)\right]^\frac{1}{2} \frac{1}{N} \sum_j \cdot \nonumber \\
& & \;\; \cdot \quad \langle
e^{-i\vec{q} ( \vec{x}_j(t) - \vec{x}_{j} )} \; D^l_{mn}(\Omega_j(t))
\; D^{l' \ast}_{m'n'}(\Omega_{j}) \rangle. 
\label{a7}
\end{eqnarray}
The reader should note that these correlation functions involve both
translational degrees of freedom {\em and} orientational degrees of freedom. They form a complete set for any two point
correlation function of an {\em arbitrary} molecular liquid and they
are also the main entities entering MMCT \cite{theisdipl}.
Specialization to $n = n' = 0$ yields the corresponding correlators
used in Ref. \cite{schilling} for {\em linear} molecules. Although
their introduction is enforced by theoretical reasons since
homogeneity and isotropy of the liquid is accounted for directly by
the Fourier transformation to the $\vec{q}$--space and their tensorial
nature, only very few of them can be measured experimentally. For
instance $S_{1mn,1mn}(\vec{q}=0,t)$ and $S^{(s)}_{2mn,2mn}(\vec{q}=0,t)$
can be obtained from dielectric measurement and NMR, respectively.
Information on the center of mass--correlator $S_{000,000}(\vec{q},t)$ follows
from light scattering, provided that the contribution of the
orientational correlators can be neglected \cite{cummins97}. The neutron
scattering cross section is a linear superposition of {\em all} of
these correlators (see e.g. \cite{theisns}) from which information on
the individual correlators could be obtained by choosing different
scattering lengths of the atomic units. On the other hand, it is a
great advantage of a MD--simulation, which really determines the
trajectories $\{\vec{x}_j(t), \Omega_j(t)\}$, that these correlation
functions can be calculated. This of course can only be done for $l$
and $l'$ smaller than a cut--off value $l_{co}$ which will 
in the analysis of the SPC/E data be chosen
as $l_{co}=2$. 

\section{Symmetry Properties}
\label{secsymmetry}
In this section we will discuss the general properties of the
correlators (\ref{a5}),(\ref{a6}) and (\ref{a7}) which follow from
symmetry. Similar discussions have been done for an expansion into
rotational invariants \cite{G&G} and in real space \cite{steele}.
The properties presented in the following will be of great importance 
for the discussion of the results of the simulation and especially
to reduce the effort for the solution of the MMCT equations which we
will present in a subsequent paper.
Concerning symmetry, we have to distinguish between {\em
global} and {\em local} symmetries where the latter are related to the
geometry of a single (rigid) molecule.

\subsection{Global Symmetry}
\label{subsecglobal}
The global symmetries arise from the invariance of the molecular
interactions under the simultaneous translation or rotation of {\em
all} molecules, provided that the external potentials are zero.
Similarly the absence of a time--dependent external force implies time
translational and time reversal symmetry which are the same as for
simple liquids and therefore will not be discussed here. The invariance
under translations in space has been already accounted for by the
transformation into $\vec{q}$--space, i.e. what remains is the
discussion of the transformation of the molecular correlators under
rotations and reflections\cite{rotation}.

Let us start with the proper transformations $R(\alpha,\beta,\gamma)
\in SO(3)$. Immediately from the transformation of
$D^l_{mn}(\Omega)$ under $R$ \cite{G&G} follows the transformation 
law for the molecular correlators:
\begin{eqnarray}
S_{lmn,l'm'n'}(\vec{q'},t) &=& \sum_{m'',m'''} \; D^{l \ast}_{m'' m}(R)
\; D^{l'}_{m''' m'}(R) \cdot \nonumber \\
& & \; \cdot \; S_{lm''n,l'm'''n'}(\vec{q},t)
\label{a8}
\end{eqnarray}
where $\vec{q'} = R \vec{q}$. The choice of the $q$--frame, i.e. 
the laboratory frame of reference oriented such that $\vec{q}$ lies
along the $z$--axis, $\vec{q} = \vec{q}_0 \equiv (0,0,q)$,  leads to a 
simplification of the molecular correlators. For a
rotation $R_z(\alpha) = R(\alpha,0,0)$ around the $z$--axis for which
$\vec{q'} = R_z \vec{q}_0 = \vec{q}_0$, eq.(\ref{a8}) implies:
\begin{equation}
\label{a9}
S_{lmn,l'm'n'}(\vec{q}_0,t) = e^{-i(m-m')\alpha} \; 
S_{lmn,l'm'n'}(\vec{q}_0,t)
\end{equation}
where $D^l_{mm'}(R_z(\alpha)) = e^{-im\alpha} \; \delta_{mm'}$
\cite{G&G} has been used. Since (\ref{a9}) is valid for {\em all}
$\alpha$, the correlators in $q$--frame must be diagonal in $m$ and
$m'$:
\begin{equation}
\label{a10}
S_{lmn,l'm'n'}(\vec{q}_0,t) = S^m_{ln,l'n'}(q,t) \; \delta_{mm'}
\end{equation}
where $q = |\vec{q}_0|$. Since (\ref{a8}) is also true for
$S^{(d)}_{lmn,l'm'n'}(\vec{q},t)$ and
$S^{(s)}_{lmn,l'm'n'}(\vec{q},t)$ the $m$--diagonality also holds
for these quantities. A further relationship follows from (\ref{a8})
by choosing a rotation $R_y(\pi) = R(0,\pi,0)$ by $\pi$ around the
$y$--axis:
\begin{equation}
\label{a11}
S_{lmn,l'm'n'}(-\vec{q}_0,t) = (-1)^{l+m+l'+m'}
S_{l\underline{m}n,l'\underline{m}'n'}(\vec{q}_0,t).
\end{equation}
Here we considered that $R_y(\pi) \vec{q}_0 = -\vec{q}_0, \;\;
D^l_{mn}(R_y(\pi)) = (-1)^{l+m} \; \delta_{m\underline{n}}$ \cite{G&G}
and $\underline{m}$ (or $\underline{n}$) denotes $-m$ (or $-n$).

Next we investigate the inversion $P$ for which $D^l_{mn}(P\Omega) =
(-1)^{l+n} \; D^l_{m\underline{n}}(\Omega)$ \cite{G&G}. Then
(\ref{a8}) yields
\begin{equation}
\label{a12}
S_{lmn,l'm'n'}(-\vec{q},t) = (-1)^{l+n+l'+n'} 
S_{lm\underline{n},l'm'\underline{n}'}(\vec{q},t).
\end{equation}
which is valid for arbitrary $\vec{q}$. 

A further useful relation is obtained taking the complex conjugate of
$S_{lmn,l'm'n'}(\vec{q},t)$ and applying
the equality $D^{l \ast}_{mn}(\Omega) = (-1)^{m+n} \;
D^l_{\underline{m} \, \underline{n}}(\Omega)$ \cite{G&G}: 
\begin{equation}   \label{eq:complex}
[S_{lmn,l'm'n'}(\vec{q},t)]^* = (-1)^{l+l'+m+m'+n+n'}
S_{l\underline{m} \, 
\underline{n},l'\underline{m}'\underline{n}'}(-\vec{q},t)
\end{equation} 
Combining eqs.(\ref{eq:complex}) respectively with eq.(\ref{a11})
and eq.(\ref{a12}) we have the final relations:
\begin{equation}
[S_{lmn,l'm'n'}(\vec{q},t)]^* = (-1)^{m+m'}
S_{l \underline{m} n,l' \underline{m}'n'}(\vec{q},t)
\end{equation} 
\begin{equation}
[S_{lmn,l'm'n'}(\vec{q},t)]^* = (-1)^{n+n'}
S_{l m \underline{n},l'm'\underline{n}'}(\vec{q},t)
\end{equation} 
or in the $q$-frame, due to the diagonality in $m$:
\begin{equation} 
[S^m_{ln,l'n'}(\vec{q}_0,t)]^* = 
S^{\underline{m}}_{ln,l'n'}(\vec{q}_0,t)
\end{equation} 
\begin{equation} \label{q2}
[S^m_{ln,l'n'}(\vec{q}_0,t)]^* = 
(-1)^{n+n'} S^m_{l\underline{n},l'\underline{n}'}(\vec{q}_0,t)
\end{equation} 
We will show in the following section that, 
in the case of molecules with 
$C_{2v}$ symmetry, like water, $n+n'$ must be even and, thus, it holds
the symmetry
\begin{equation} \label{q3}
[S^m_{ln,l'n'}(\vec{q}_0,t)]^* = 
S^{m}_{l\underline{n},l'\underline{n}'}(\vec{q}_0,t)
\end{equation} 

%A further property not related
%to symmetry follows from the definitions eqs. (\ref{a5})-(\ref{a7}):
%\begin{equation}
%\label{a13}
%S^\ast_{lmn,l'm'n'}(\vec{q},t) = S_{l'm'n',lmn}(\vec{q},t)
%\end{equation}
%and  using $D^{l \ast}_{mn}(\Omega) = (-1)^{m+n} \;
%D^l_{\underline{m}\underline{n}}(\Omega)$ \cite{G&G} we get:
%\begin{eqnarray}
%S^\ast_{lmn,l'm'n'}(\vec{q},t) &=& (-1)^{l+m+n+l'+m'+n'} \cdot 
%\nonumber \\ & & \; \cdot S_{lmn,l'm'n'} (-\vec{q},t).
%\label{a14}
%\end{eqnarray}
%Eq.(\ref{a13}) means that the infinite dimensional correlation matrix
%is hermitian. We stress that in general eqs.(\ref{a11})-(\ref{a14}) do
%{\em not} imply that the correlators are real in the $q$--frame, in
%contrast to linear molecules \cite{dumbbell,schilling}.

%Eq. \ref{a13} and Eq.\ref{a15} are also valid in the $q$-frame.
%Thus in the $q$-frame the following properties hold  
%for all correlators 
%(both self and distinct), independently from the molecular shape:

%\begin{itemize}

%\item{} m-diagonality, Eq. \ref{a11}

%\item{} $m  -> \underline{m}$, from Eq.\ref{a12} and Eq. \ref{a14} 
%\begin{equation}
%\label{q2}
%S^\ast_{lmn,l'm n'}(\vec{q_0},t) = (-1)^{n+n'} 
%S_{l\underline{m}n,l'\underline{m}n'}(\vec{q_0},t)
%\end{equation}  
%In the case of molecules with $C_{2v}$ symmetry, like water, 
%$n+n'$ must be even, as shown in the next subsection.

%\item{} $n,n'  -> \underline{n},\underline{n}'$, 
%from from Eq.\ref{a13} 
%and Eq.\ref{a14} 
%\begin{equation}
%\label{q3}
%S^\ast_{lmn,l'm n'}(\vec{q_0},t) =  
%S_{lm\underline{n},lm'\underline{n}'}(\vec{q_0},t)
%\end{equation} 

%\end{itemize}

\subsection{Local Symmetry}
\label{subseclocal}
In case that a molecule has a point symmetry we can derive additional
identities for $S_{lmn,l'm'n'}(\vec{q},t)$. Since we have water in
mind which has a $C_{2v}$--symmetry we will discuss molecules 
with $C_{rv}$ symmetry (see e.g. \cite{A&M}). To avoid
confusion with the $n$--index of the rotation matrices $D^l_{mn}$ we
deviate from the conventional notation $C_{nv}$. This symmetry means
that the molecule possesses a $r$--fold rotational axis and $r$ planes
of reflection symmetry which contain the rotational axis. It is
obvious that application of the local symmetry operations to any
single molecule must leave the interaction invariant. Let us begin
with the $r$--fold rotational symmetry. Without restriction of
generality we choose the body--fixed $z$--axes along the $r$--fold
symmetry axis. In that case the $r$--fold symmetry only affects the
third Euler angle $\chi$. The transformation
\begin{equation}
\label{a15}
\chi_j \to \chi_j + \nu \frac{2\pi}{r}
\end{equation}
for $\nu$ an integer and {\em fixed} $j$ implies that
\begin{equation}
\label{a16}
\chi_j(t) \to \chi_j(t) + \nu \frac{2\pi}{r}.
\end{equation}
Since $D^l_{mn}(\Omega_j(t)) = e^{-im\phi_j(t)} \;
d^l_{mn}(\theta_j(t)) \; e^{-in\chi_j(t)}$ \cite{G&G} it follows with
the {\em separate} use of (\ref{a15}) and (\ref{a16}) to the {\em
distinct part} (\ref{a6}):
\begin{equation}
\label{a17}
e^{i\nu\frac{2\pi}{r}n} = 1 \quad \mbox{and} \quad
e^{-\nu\frac{2\pi}{r}n'} = 1
\end{equation}
for all integers $\nu$. This restricts $n$ and $n'$ to integer
multiples of $r$, i.e.
\begin{equation}
\label{a18}
S^{(d)}_{lmn,l'm'n'}(\vec{q},t) = \left\{ 
\begin{array}{c@{\quad}c}
\not= 0 & \mbox{for} \; n,n' \in \{0,\pm r, \pm 2r, ... \} \\
0 & \mbox{otherwise} 
\end{array}
\right.
\end{equation}
For the {\em self part} (eq.(\ref{a7})) we have to use (\ref{a15})
{\em and} (\ref{a16}) {\em simultaneously}, leading to:
\begin{equation}
\label{a19}
e^{i\nu\frac{2\pi}{r}(n-n')} = 1
\end{equation}
for all integers $\nu$, which is fulfilled for $n' = n (mod \; r)$, i.e.
\begin{equation}
\label{a20}
S^{(s)}_{lmn,l'm'n'}(\vec{q},t) = \left\{ 
\begin{array}{c@{\quad}c}
\not= 0 & \mbox{for} \; (n-n') \in \{0,\pm r, \pm 2r, ... \} \\
0 & \mbox{otherwise} 
\end{array}
\right.
\end{equation}

According to these results the total correlator
$S_{lmn,l'm'n'}(\vec{q},t)$ (eq.(\ref{a5})) reduces to its
self part for $n$ or $n'$ not equal to an integer multiple of $r$. 
For the case of water (r=2), eqs.(\ref{a19}) 
requires $n+n'$ to be even, a condition which simplifies  
eq.(\ref{q2}) further.

What remains to be discussed is the role of the reflection symmetry.
Here we restrict ourselves to the {\em static} correlators. It will be
shown in Appendix \ref{appsym} that for their distinct part it
implies:
\begin{equation}
\label{a21}
S^{(d)}_{lmn,l'm'n'}(\vec{q}) \to  
S^{(d)}_{lm|n|,l'm'|n'|}(\vec{q}) \quad , \; T \to \infty
\; ,
\end{equation}
i.e. it does not depend on the sign of $n$ and $n'$.
%In case that e.g. $m' \not= 0$ and $l'$ odd, 
%but $m = 0$ or $l$ even eq.
%(\ref{a21}) is modified to:
%\begin{equation}
%\label{a21'}
%S^{(d)}_{lmn,l'm'n'}(\vec{q}) = S^{(d)}_{lm|n|,l'm'n'}(\vec{q}),
%\end{equation}
%i.e. the distinct part depends on the sign of $n'$ but not on that of
%$n$. A similar identity follows, if $m \not= 0$ and $l$ odd but $m' =
%0$ or $l'$ even. In that case $S^{(d)}_{lmn,l'm'n'}(\vec{q})$ does not
%depend on the sign of $n'$. Note that in the $q$--frame $m$ equals
%$m'$. Therefore $m = 0$ implies $m' = 0$ such that (\ref{a21}) holds.

We close this section by a discussion of the implications following
from the results in both subsections. It will be crucial that we will
consider the correlators in the $q$--frame. The correlators in an
arbitrary reference frame are easily obtained from eq.(\ref{a8}).
Since the identities derived in subsection \ref{subsecglobal} also
hold for $S^{(s)}_{lmn,l'm'n'}(\vec{q},t)$ and
$S^{(d)}_{lmn,l'm'n'}(\vec{q},t)$ we get from
eqs.(\ref{a10})-(\ref{eq:complex}) with (\ref{a18}) and (\ref{a21}) 
for {\em static} correlators in case of high temperatures:
\begin{equation}
\label{a22}
(S^{(d) m}_{ln,l'n'}(q))^\ast \cong S^{(d) m}_{ln,l'n'}(q) \cong S^{(d)
\underline{m}}_{ln,l'n'}(q),
\end{equation}
i.e. the static distinct part is real and does not depend on the sign
of $m$. Since the static self part fulfills:
\begin{equation}
\label{a23}
S^{(s) m}_{ln,l'n'}(q) = \delta_{ll'} \; \delta_{nn'}
\end{equation}
it follows from (\ref{a5}), (\ref{a21}) and (\ref{a23}) for high
temperatures that
\begin{equation}
\label{a24}
S^m_{ln,l'n'}(q) \cong \left\{
\begin{array}{c@{\;\; , \;\;}c}
S^{(d) m}_{l|n|,l'|n'|}(q) & n \not= n' \\
1 + S^{(d) m}_{l|n|,l'|n'|}(q) & n = n'
\end{array}
\right.
\end{equation}
i.e. $S^m_{ln,l'n'}(q)$ is approximately determined by $S^{(d)
m}_{l|n|,l'|n'|}(q)$ and in addition it is real and does not depend on
the sign of $m$, due to (\ref{a22}).

%is
%the position of the center of mass of the $i$th molecule and 
%$\Omega_i=(\phi_i,\theta_i,\chi_i)$ are the Euler angles of the same 
%molecule. A statistical description of the 
%behavior of the liquid is obtained 
%studying the correlation functions of the density fluctuations.
%It is useful to perform an expansion of the microscopic density 
%on the basis of the generalized spherical harmonics $D^l_{mn}(\Omega)$
%\cite{G&G}, where the indexes $l,m,n$ are integer 
%with $l \ge 0$, $m=-l,l$ and $n=-l,l$. 
%Thus, in the Fourier space, the rotational
%correlation functions are
%\begin{equation}   
%S_{lmn,l'm'n'}(\vec{q},t) \equiv {1 \over N} 
%<\rho_{lmn}^*(\vec{q},t) \rho_{l'm'n'}(\vec{q})>
%\label{eq:sq}
%\end{equation}
%where
%\begin{equation}
%\rho_{lmn}(\vec{q},t)=(i)^l \sqrt{(2l+1)} \sum_i e^{i \vec{q} \cdot 
%\vec{x_i(t)}}  D^l_{mn}(\Omega_i(t))^*
%\label{eq:rho}
%\end{equation}
%The prefactor in eq.(\ref{eq:rho}) is chosen in 
%such a way that the asymptotic
%value for the diagonal correlators ($l=l'$, $m=m'$ and $n=n'$) is one.
%With this choice, for a liquid of 
%linear ({\bf Mainz group: planar ????? 
%please  confirm !})
%molecules the static structure factors are real.
%For $l=0$ (which also requires $m=n=0$) 
%eq.(\ref{eq:sq}) defines the center of mass
%structure factor, which coincides with the static $S_q$ 
%defined for simple liquids.

\section{Results}
\label{sec:res}
In this section we present a detailed analysis of the static and
dynamic rotational correlators as calculated from a Molecular Dynamics
(MD) simulation of a molecular network-forming liquid. The system is a
liquid of 216 rigid molecules whose geometric
parameters are chosen in such a way to mimic water molecules. The
intermolecular interactions are described by the SPC/E (Single Point
Charge Extended) potential\cite{spce} which has been shown to
be able to describe most of the thermodynamic properties of 
supercooled liquid water. We will not report here the
details of the simulation which can be found in 
refs.\cite{self,collective}.

The low T simulations have been run for more than $50^8$ integration
timesteps, corresponding to 50 ns. Each of the simulation runs requests
several months of CPU time  on one 
400 MHz alpha-processor for equilibration and
more than one year of computer time 
to generate the configuration ensemble studied. In this situation, 
equilibrium runs for much larger systems at low temperature are 
very time consuming. 
Since periodic boundary conditions
may introduce distortions in angular properties near the boundaries, since ---
as it will be shown in the following --- some of the molecular correlators
present unexpected peaks at wave vectors smaller than the center-of-mass peak
position, corresponding to a distance of about three molecular diameters, and
since the amplitude of these peaks grows at low temperature, we have made the
additional effort to equilibrate a system made of 1728 molecules at $T=207 K$
to make sure that no finite size effects show up in the molecular correlators
for the system size considered.

In order to calculate the Euler angles of each molecule the 
molecular reference frame has been chosen in such a way that the
$z$-axes of the body--fixed frame has the same direction as
the dipole of the molecule and the $y$-axes
lies along the line connecting the two hydrogens. 
The $x$-axes is therefore
determined to be orthogonal to the molecular plane. 
With this choice of the molecular axes the structure factors 
eq.(\ref{eq:sq}) 
reduce, in the limit $|\vec{q}| \to 0$ and for $n=n'=0$, to
\begin{equation}
S_{lm0,l'm'0}(0,t) = C_l(t) \delta_{ll'}\delta_{mm'}
\end{equation}
where 
\begin{equation}
C_l(t) = {1 \over N} \sum_{i,j} \langle P_l(\vec{e}_i(t) \cdot 
\vec{e}_j(0))
\rangle
\label{eq:cl}
\end{equation}
are the orientational correlation functions of the dipolar moments 
$\vec{\mu}_i(t) = \mu \vec{e}_i(t)$.
In eq.(\ref{eq:cl}) $P_l$ is the Legendre polynomial of order $l$ and
$\mu$ is the dipole strength. These 
$q$-independent rotational correlation functions can be 
experimentally measured
for some values of $l$ (see above). 
A complete analysis of the self part of
$C_l(t)$ for the system under investigation is 
treated in ref.\cite{vigo}.
In what follows the discussion will be extended to the 
generalized correlators $S^m_{ln,l'n'}(q,t)$ in the $q$--frame.

In the first subsection we will discuss the static correlators and in
the second subsection we will turn to the detailed analysis of the 
dynamic behavior of the correlation functions defined in 
eq.(\ref{eq:sq}) as calculated from the MD data. We will discuss the
numerical results comparing them to the qualitative and quantitative 
predictions of MCT or MMCT.

\subsection{Static properties}

We have calculated all the static structure factors as defined in 
eq.(\ref{eq:sq}) up to  $l=2$. 
The $q$-vector range has been chosen in such a 
way to include all the meaningful structure of 
each correlation function, i.e. 
the value of $q$ ranges from $3.3 nm^{-1}$, 
which is the lower bound imposed by the finite size, up to around 
$110 nm^{-1}$, a value at which all the 
structure factors have practically 
reached their asymptotic values. 
The grid spacing has been fixed to $\Delta q=1.11 nm^{-1}$
which allows all the peaks in the correlators to be well resolved.   

\begin{figure}
\epsfxsize=7.5cm
\epsfysize=15cm
\centerline{\epsffile{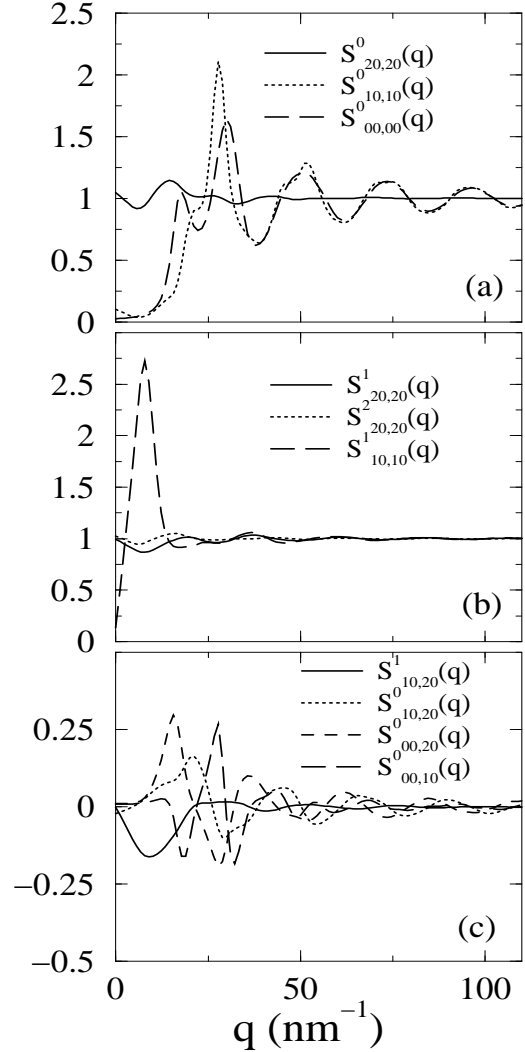}}
\caption{Static structure factors $S^m_{ln,l'n'}(q)$. 
Diagonal correlation functions with $n=n'=0$ are shown in (a) and (b). 
Off-diagonal correlation functions with $n=n'=0$ are shown in (c).
}
\label{fig:sqst1}
\end{figure}

\begin{figure}
\epsfxsize=7.5cm
\epsfysize=15cm
\centerline{\epsffile{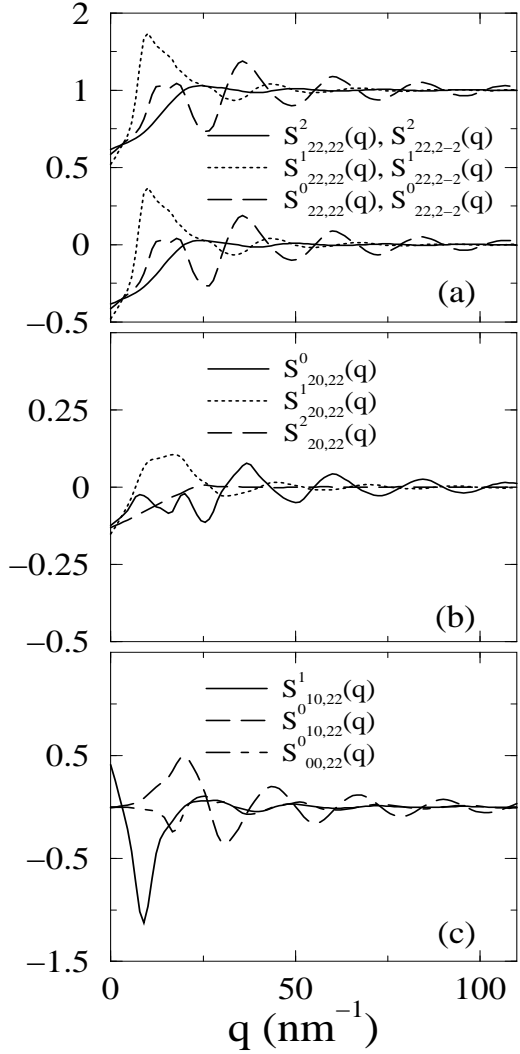}}
\caption{Static structure factors 
$S^m_{ln,l'n'}(q)$ with $n$ and/or $n' \ne 0$.
Subfigure (a) demonstrates the 
validity of eq.(\protect\ref{a24}).The three 
upper lines refer to the 
case $n=n'$ while the lower ones represent $n=-n'$.}
\label{fig:sqst2}
\end{figure}

In order to obtain a more significant statistics, the static structure
factors have been calculated, 
for each configuration, for several directions
of the $q$-vectors with respect to the fixed 
laboratory frame. The different 
results are averaged after performing a suitable rotation which
brings the $q$-vector along the $z$-axes of the lab frame  
(eq.(\ref{a8})). Furthermore the resulting correlation 
functions have been averaged
over several configurations extracted from 
a time evolution which extends
up to $250 ns$. The global symmetries described in 
section \ref{secsymmetry},
which are all fulfilled within the numerical error by the
numerical correlators, allow 
a further average of the static correlators. 
In this way the numerical noise is reduced to its
lowest  possible value.
The resulting structure factors are shown in 
Figs. \ref{fig:sqst1}  and \ref{fig:sqst2}
for the lowest simulated temperature $T=207~K$.
We found that within the numerical error, all correlators are real.

\begin{figure}
\epsfxsize=7.5cm
\epsfysize=15cm
\centerline{\epsffile{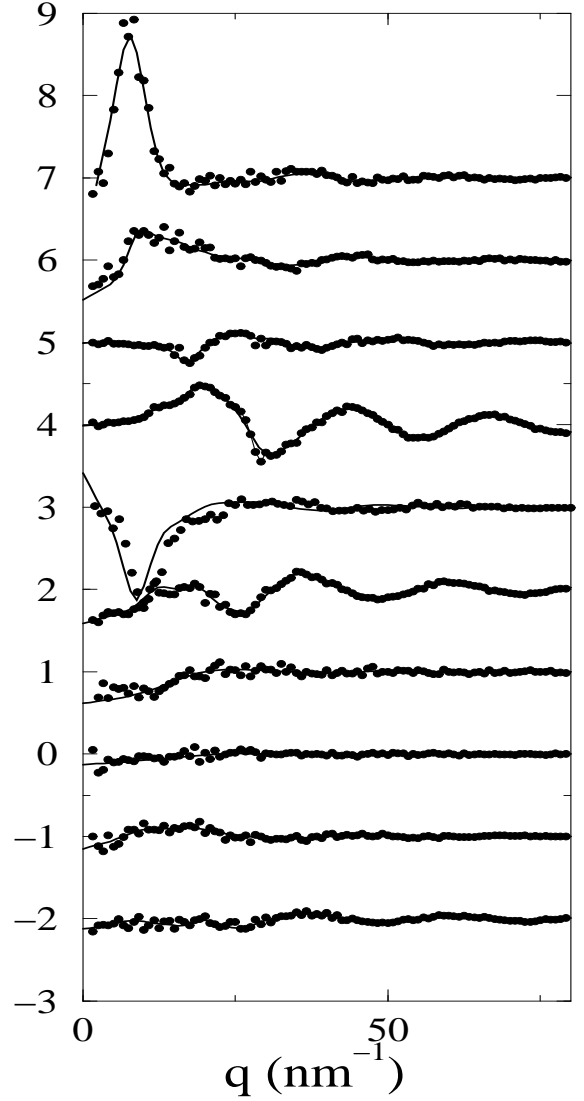}}
\caption{Comparison of the static structure factors $S_{ln,l'n'}^m(q)$ at
$T=207K$ between the $6^3$ (solid lines) and the $12^3$ (symbols) molecules 
system. 
The comparison is presented for several $S_{ln,l'n'}^m(q)$, which have
been arbitrarily shifted along the $y$-axis to improve the quality of the
figure. From top to bottom:$S_{10,10}^1$, $S_{22,22}^1$, $S_{00,22}^0$,
$S_{10,22}^0$, $S_{10,22}^1$, $S_{22,22}^0$, $S_{22,22}^2$, $S_{20,22}^2$, 
$S_{20,22}^1$, $S_{20,22}^0$. 
The molecular correlators which have a peak at small $q$ vectors
are shown, to highlight the absence of finite size effects even in the worst
case where the molecular correlation extends over about three molecular
diameters.}
\label{fig:fscheck}
\end{figure}

Some of the static correlators shown in Figs. \ref{fig:sqst1} and
\ref{fig:sqst2} are characterized by large amplitude peaks at $q$-vectors
corresponding to distances of the order of three molecular diameters. This is
exactly the distance where finite size effects could be expected for a system
composed of $6^3$ molecules. To check if the results are real or simply
artifacts of the simulation conditions, we report in Fig. \ref{fig:fscheck}
the static structure factors for the $6^3$ and for the $12^3$ molecule system.
We perform the check at the lowest studied temperature, where the angular
correlations are enhanced. Although the statistic for the large system is
poorer, as expected from the shorter simulation time, both position and
amplitude of the peaks are not affected. Data in Fig \ref{fig:fscheck}
confirms that the angular correlations do persist longer than the center of
mass correlations and that the static molecular correlators, i.e. the
quantities which are requested as input by the molecular mode coupling theory,
are not affected by the size of the studied system.

The different figures list the static correlation functions 
in terms of progressive
angular complexity. Fig.\ref{fig:sqst1}(a,b) includes all the diagonal
correlators (i.e.  $l=l'$) with $n=n'=0$. 
In Fig. \ref{fig:sqst1}(c)
$n$ and $n'$ are still fixed to zero but 
the diagonality in $l$ and $l'$ is 
now relaxed. In Fig.\ref{fig:sqst2} the correlators with 
$n$ and/or $n'$ different from zero are reported.
Figs.~\ref{fig:sqst1} and  \ref{fig:sqst2} show that several molecular 
correlators are comparable in intensity 
to the center of mass correlation function. 
Furthermore, some of them
present completely new features, i.e. the $q$-dependence of the 
molecular correlations with $l$ and $l'$ different from zero is 
completely different from the center of mass structure 
factor.
Some of the  $S^m_{ln,l'n'}(q)$ display 
peaks and minima at  $q$-vector values where 
the center of mass structure factor is rather structureless. 
The generalized spherical harmonics $D^l_{mn}(\Omega_j)$ in 
eqs.(\ref{a6}) and (\ref{a7}) can be viewed as weights that "single
out" molecules with certain orientations $\Omega_j$. The differences
in the $q$--dependence of the generalized orientational correlators
show that looking at molecules with "selected" orientation reveals
characteristic length scales of the system which are different from the
center of mass ones. This can result in the shift of a peak as in the case of the
main peak of $S^0_{10,10}(q)$ compared to $S^0_{00,00}(q)$ (see
Fig.~\ref{fig:sqst1}(a)). The "proper choice" of orientations can also
reveal order on scales that are longer than the typical intermolecular
distance and that is not visible in the center of mass correlators. For example,
the most significant peak in $S^1_{10,20}(q)$ and $S^1_{10,10}(q)$
is located around $q=6.5 nm^{-1}$, a $q$-vector much smaller than the 
center of mass structure factor first peak.
The $q=6.5 nm^{-1}$ peak, which appears 
in correlators with $l$ or $l'$ equal to one and $n=n'=0$, 
may reflects the strong dipolar interactions
characteristic of water. At distances of the order of $2\pi/6.5$ nm,
about three molecular diameters,  
the SPC/E potential is equivalent to the potential generated by 
an electrostatic dipole with the same dipole moment of the SPC/E 
model. 
Thus, data in Fig.~\ref{fig:sqst1}, \ref{fig:sqst2}
suggest that
in water angular correlations persist over 
distances much longer than the 
center of mass correlations. 
We also point out that the correlators in 
Fig.~\ref{fig:sqst1}(c) have an intensity more than three times smaller 
than the  ones reported in 
Fig.~\ref{fig:sqst1}(a) (where for the latter 
one has to use as intensity
$S^m_{l0,l0}(q) - 1$). This
suggests that the diagonal static correlation function  
could provide a good starting approximation for a 
MMCT description of the slowing down of dynamics in
SPC/E water. We recall that this is at odd with  the case of a 
liquid of Lennard Jones dumbbells\cite{kks} where 
the off-diagonal terms have a large amplitude.  
Correlators with $n$ different from zero (see Fig. \ref{fig:sqst2})
carry information about the planar shape of the molecule 
($n=n'=0$ is equivalent to assuming that the Euler angle $\chi$ 
is always zero), i.e. on the absence of cylindrical symmetry. 
Subfigures (a) of that figure demonstrates the validity of the
property (\ref{a24}).
Data in Fig. \ref{fig:sqst2} shows that a few of these correlators
describe a significant amount of angular correlations. Again, the 
largest amplitude is observed for the case in 
which $l$ or $l'$ is one, i.e.
for $S^1_{10,22}(q)$.

Figure \ref{fig:sqst2} also exhibits the $n \to -n$ (or
$n' \to -n'$) symmetry as given by eq.(\ref{a24}).
Although this symmetry becomes exact at least at high temperatures (see
Appendix \ref{appsym}), it seems to be valid within the numerical error
even at $T = 207 K$, which is not high anymore. Although in Appendix
\ref{appsym} it is shown that for {\em any} temperature contributions
to the distinct part exist for which this symmetry is still exact, it
is not obvious to us why their weight is so large.

\begin{figure}
\epsfxsize=7.5cm
\epsfysize=15cm
\centerline{\epsffile{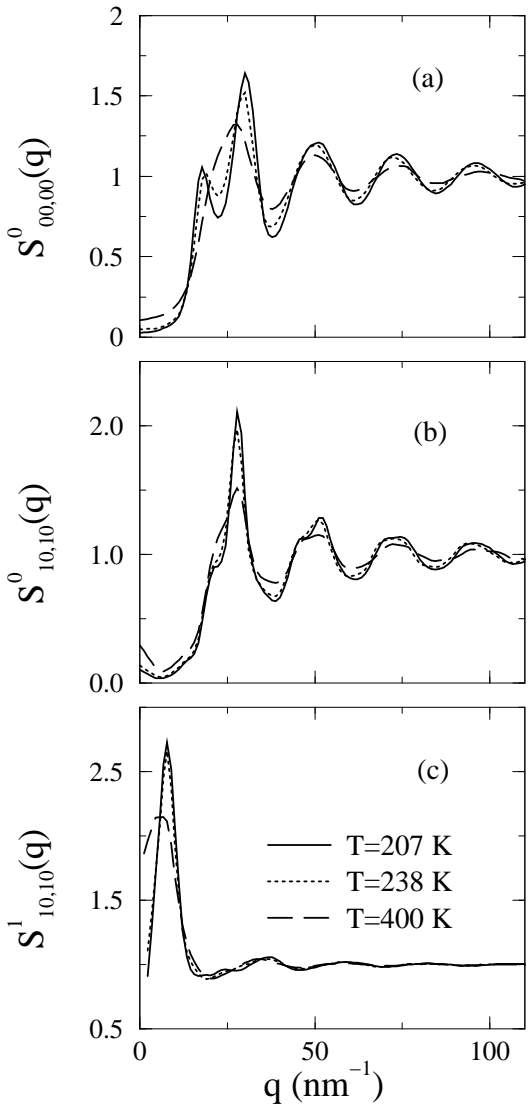}}
\caption{Temperature dependence of the diagonal 
static structure factors  with $l=0,1$ and $n=n'=0$. 
}
\label{fig:sqT-fig1}
\end{figure}

\begin{figure}
\epsfxsize=7.5cm
\epsfysize=15cm
\centerline{\epsffile{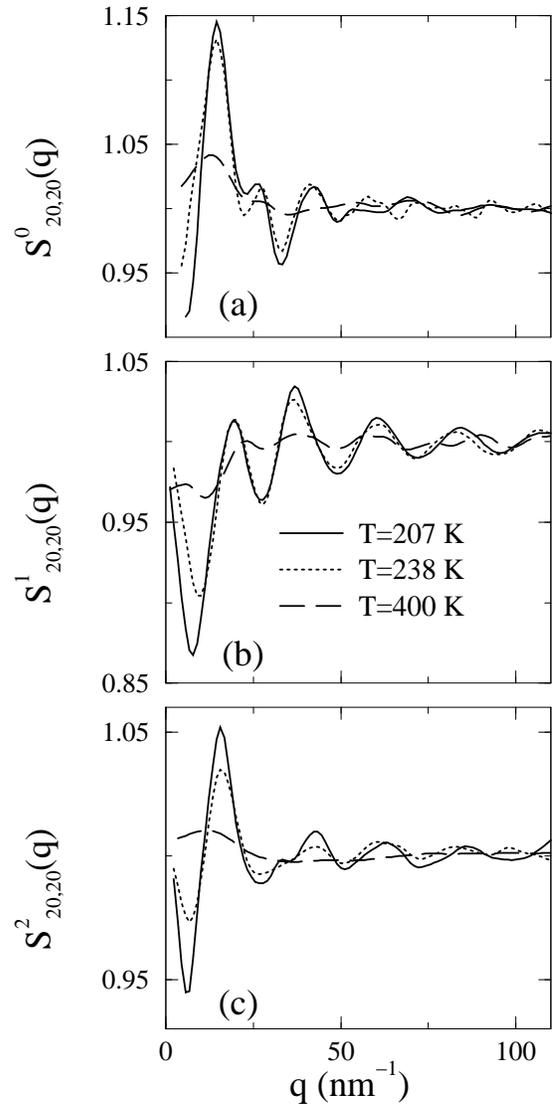}}
\caption{Temperature dependence of the diagonal 
static structure factor  with $l=2$ and $n=n'=0$.  
}
\label{fig:sqT-fig2}
\end{figure}

\begin{figure}
\epsfxsize=12cm
\epsfysize=10cm
\centerline{\epsffile{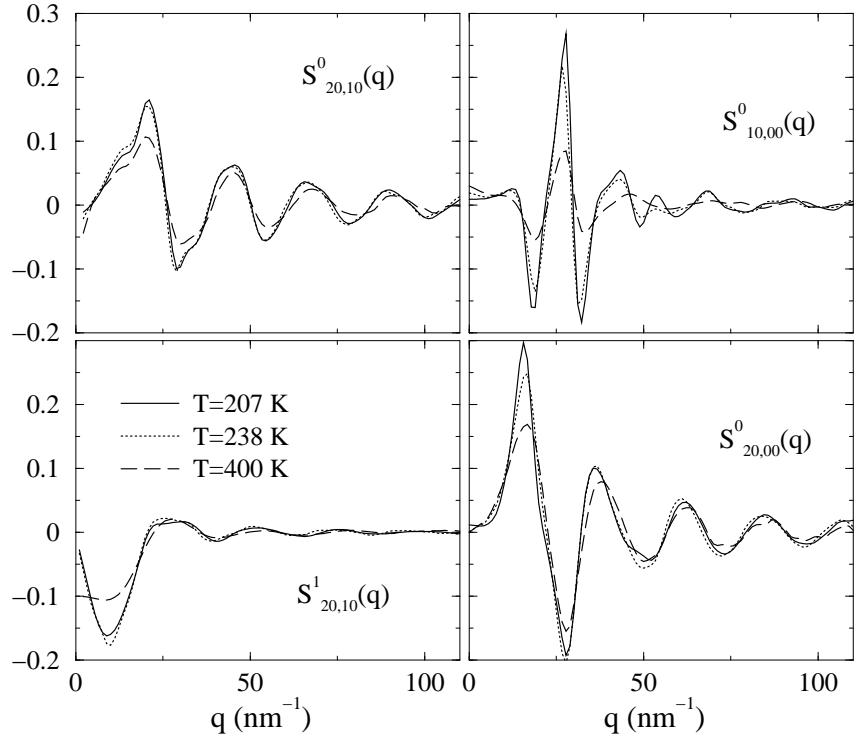}}
\caption{Temperature dependence of the off-diagonal 
static structure factor  with  $n=n'=0$.  
}
\label{fig:sqT-nodiag}
\end{figure}

\begin{figure}
\epsfxsize=12cm
\epsfysize=10cm
\centerline{\epsffile{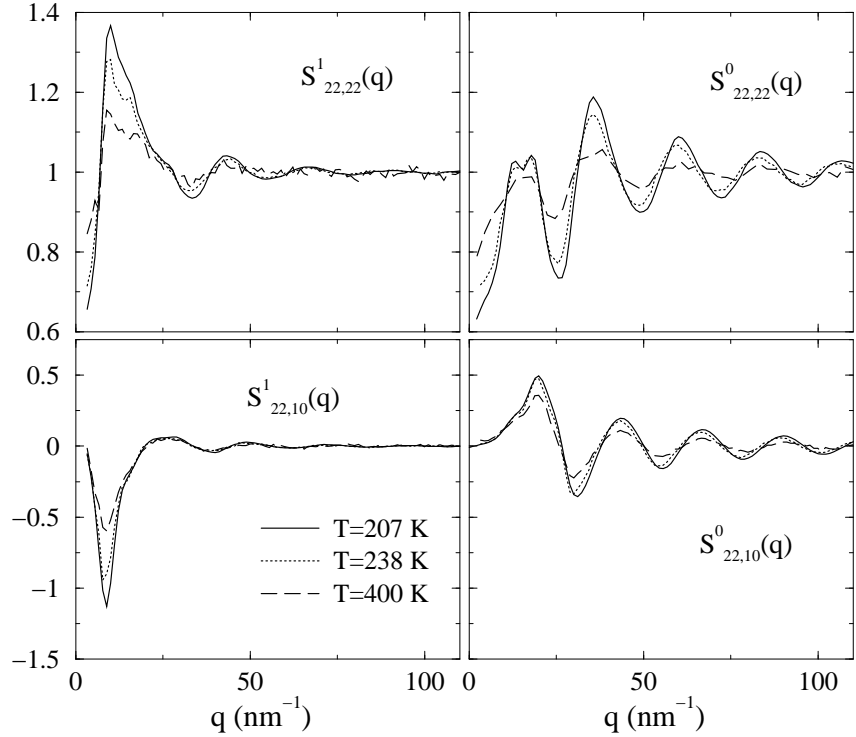}}
\caption{Temperature dependence of some 
static structure factor  with  $n$ and/or $n' \ne 0$.  
}
\label{fig:sqT-due}
\end{figure}

Figs.\ref{fig:sqT-fig1},~\ref{fig:sqT-fig2}, 
~\ref{fig:sqT-nodiag} and  \ref{fig:sqT-due} 
show the temperature dependence of 
the static structure factors $S^m_{ln,l'n'}(q)$.
In order to analyze
the relative variation of the correlators on varying $T$ we report 
three different temperatures, i.e. $T=207~K$, $T=238~K$ and
$T=400~K$.
Figs.~\ref{fig:sqT-fig1} and \ref{fig:sqT-fig2} report the 
diagonal  $S^m_{l0,l0}(q)$. Fig.~\ref{fig:sqT-nodiag} reports the 
off diagonal terms with $n=n'=0$, while  Fig.~\ref{fig:sqT-due} 
reports some correlation functions with $n \ne 0$.
From these figures we see that the behavior of 
$S^m_{ln,l'n'}(q)$ on changing $T$ is strongly correlator dependent. 
The static structure factor $S^0_{00,00}(q)$ (center of mass) 
shows a significant increase of the 
resolution of the peaks, which become sharper as $T$ is lowered, 
especially
at small $q$-vectors. 
The diagonal correlators with $l=l'=1$ are less temperature dependent. 
They remain almost unchanged 
at low temperatures changing $T$ by 
$30$ degrees, from $T=238~K$ to $T=207~K$, i.e. in the region where the 
molecular diffusivity decreases by more than 2 
order of magnitude\cite{self}.
Thus, while these correlators have a large amplitude, which implies
that they might contribute substantially to the MMCT vertices,
the dynamical transition may not be  controlled by them.
The off-diagonal terms shown in Fig. \ref{fig:sqT-nodiag} as well as the
static correlation functions with $n$ and/or $n' \ne 0$ 
(Fig.~\ref{fig:sqT-due}) also do not show significant temperature 
variation at low temperatures.
This suggests that a small set of diagonal correlators may 
play the relevant role in the slowing down of the molecular
dynamics on supercooling. Of course, only a  full MMCT calculation 
can confirm such hypothesis and can provide definitive answers on 
which modifications of the angular correlations drive the
ideal glass transition.
\onecolumn

\twocolumn

\subsection{Dynamic properties}

In this section we discuss the time evolution of
the molecular structure factors with $l,l' \le 2$ at 
different $q$-vector 
values.  We present the results of the calculation
in the perspective of testing how far the mode coupling framework can 
describe the behavior of $S^m_{ln,l'n'}(q,t)$. 
Since the complete MMCT equations have not yet been
solved for the complete dynamic evolution of the correlators, we mostly
keep our discussion to a qualitative level. We  show that most of
the universal predictions (see Introduction) for the slow 
relaxation in supercooled liquids
are excellently satisfied by the numerical data. 

One of the strongest predictions of MCT is the validity of 
the so called 
time-temperature superposition principle in the 
$\alpha$-relaxation region (see Introduction). 
The time-temperature superposition principle states that in a wide range of 
temperatures above the MCT critical point (which will be defined in the 
following) it is possible to scale the same correlator evaluated at 
different $T$ on a single master curve through a 
rescaling of the time, i.e.

\begin{equation}
\phi(t)=\tilde{\phi}(t/\tau(T))
\label{eq:master}
\end{equation}

In eq.(\ref{eq:master}) $\phi(t)$ indicates any dynamic structure
factor, $\tilde{\phi}(\tilde{t})$ the master function and $\tau(T)$ 
is a temperature dependent 
time scale which is characteristic of the chosen correlator. 
The temperature dependence of the time scale is also predicted by MCT
or MMCT.
In leading order $\tau(T)$ is a power
law diverging at the critical transition temperature:
 
\begin{equation}
\tau(T) \sim |T-T_c|^{-\gamma}
\label{eq:gamma}
\end{equation}

\begin{figure}
\epsfxsize=12.5cm
\epsfysize=10cm
\centerline{\epsffile{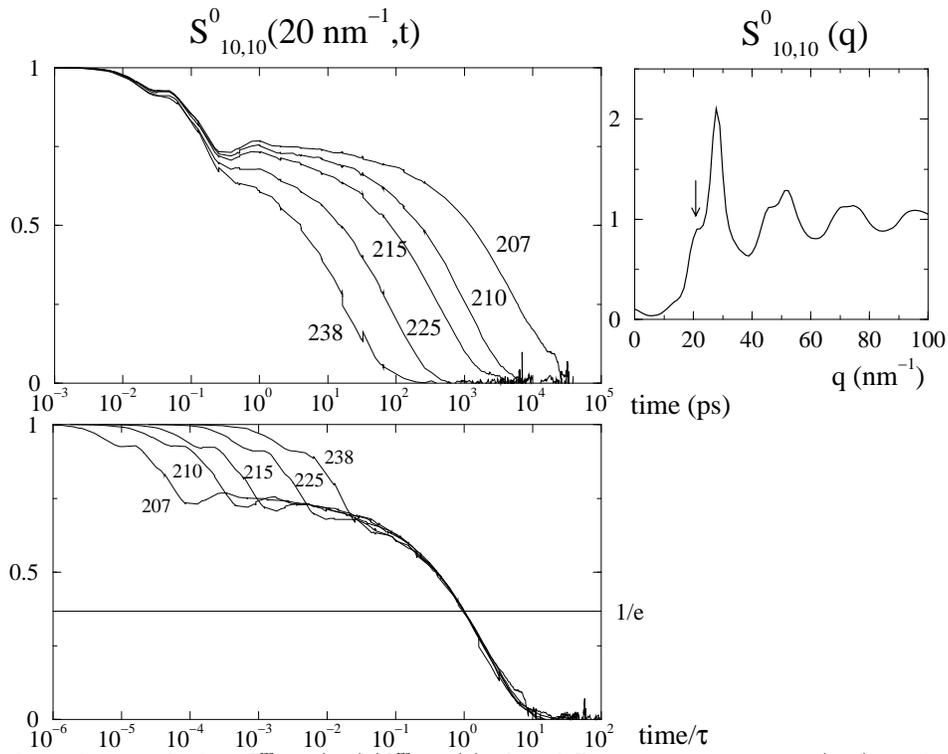}}
\caption{Time dependence of the $S^m_{ln,l'n'}(q,t)/S^m_{ln,l'n'}(q)$ 
for different temperatures (top) and the corresponding 
time/temperature scaling representation (bottom). 
The small inset shows the corresponding static structure factor 
with an indication of the chosen $q$-vector (see arrow). 
This figure refers to 
 $S^0_{10,10} (q=20 nm^{-1},t)$. }
\label{fig:ttspUZZUZZ1}
\end{figure}

\begin{figure}
\epsfxsize=12.5cm
\epsfysize=10cm
\centerline{\epsffile{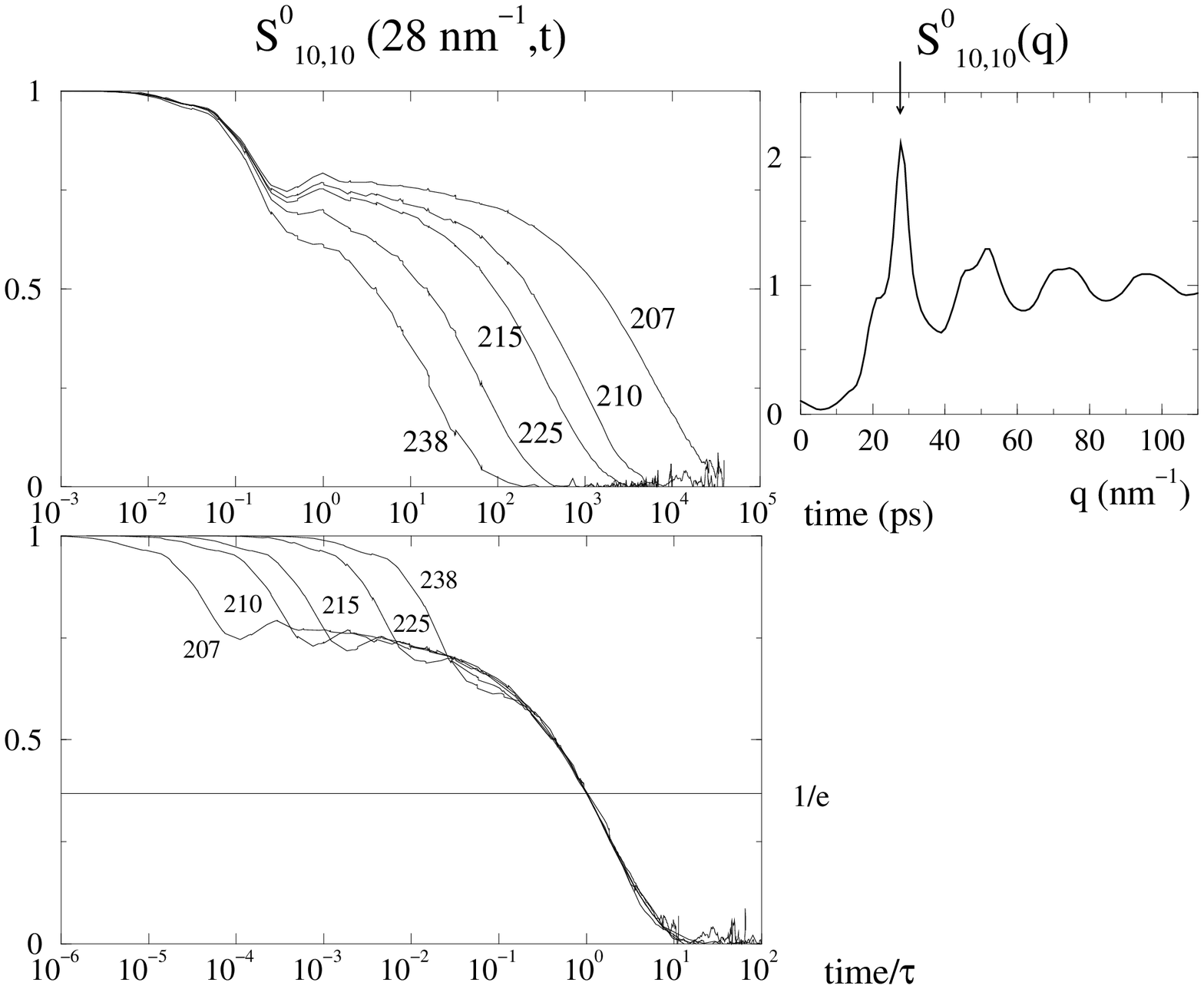}}
\caption{Same as \protect\ref{fig:ttspUZZUZZ1} 
for  $S^0_{10,10}(q=28 nm^{-1},t)$.
}
\label{fig:ttspUZZUZZ2}
\end{figure}

\begin{figure}
\epsfxsize=12.5cm
\epsfysize=10cm
\centerline{\epsffile{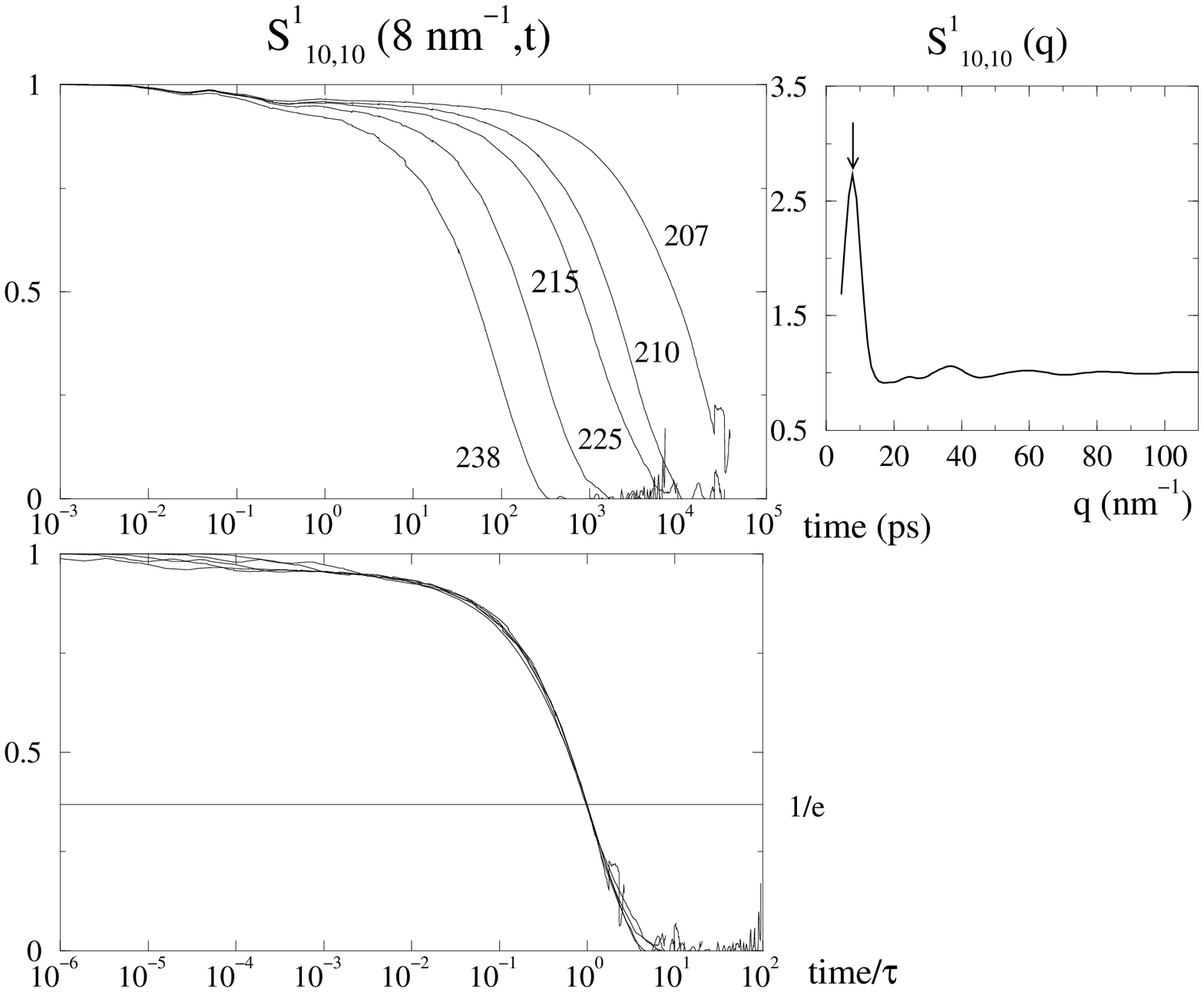}}
\caption{Same as \protect\ref{fig:ttspUZZUZZ1} 
for  $S^1_{10,10}(q=8 nm^{-1},t)$.
}
\label{fig:ttspUUZUUZ1}
\end{figure}

Eq.(\ref{eq:gamma}) is one of the possible operative definitions of the 
mode-coupling critical temperature $T_c$. $T_c$ is 
the temperature at which the characteristic time scales diverge, i.e the
point at which the dynamic of the liquid is completely frozen. 
Thus $T_c$
defines a kinetic transition from an 
ergodic to a nonergodic dynamic. This
is a purely kinetic transition which 
does not have a thermodynamic counterpart.
In real liquids no sharp transition is observed and 
close to $T_c$ the system switches to a different dynamic where 
hopping phenomena become dominant. Thus $T_c$ assumes the meaning of a 
crossover temperature. 
The transition temperature $T_c$ and 
the scaling exponent $\gamma$ are not 
universal quantities since they strongly 
depend on the physical system under
investigation and on the volume and pressure conditions, but they have
a sort of ``universality'' in the sense that they are 
predicted to be correlator independent. Indeed, $T_c$ and $\gamma$ 
are predicted to have the 
same value for all correlation functions which couple to each other.
In case of molecular liquids their values can be obtained from MMCT.
In previous papers\cite{self,collective,vigo} it has been shown that
in the case of SPC/E, the
time-temperature superposition principle is satisfied by self and collective center of mass correlators and also by
the $q$-independent rotational correlation functions. It has been 
also shown that $T_c=200 \pm3 $   
and the critical exponent $\gamma=2.7$ are the
same for all examined correlators within the numerical error.
Here we generalize
this conclusion showing that the time-temperature superposition principle also holds for all the molecular 
collective  correlators
up to $l=2$. We  show here only two representative correlators.
In Fig. \ref{fig:ttspUZZUZZ1} we have reported  
$S^0_{10,10}(q,t)$ at the different
temperatures normalized to its static value $S^0_{10,10}(q)$. We have
rescaled each curve choosing as $\tau(T)$ the time at which the
correlator has decayed to the value $1/e$. 
The different curves perfectly
overlap in the $\alpha$-region confirming the validity of the time-temperature superposition principle. 
The small graph shows as a reference 
the static structure factor; an arrow 
is pointing to the $q$-value for which the analysis is performed. 
In Fig. \ref{fig:ttspUZZUZZ2} 
we show the test of the time-temperature superposition principle for the 
same correlator but for a different 
value of $q$,
while Fig. \ref{fig:ttspUUZUUZ1} shows the same analysis for 
a different correlator, $S^1_{10,10}(q,t)$. These results, and similar 
analysis for the other correlators (not reported in this Article) lead
to the conclusion that the time-temperature superposition principle is 
satisfied for all examined $S^m_{ln,l'n'}(q)$.

\begin{figure}
\epsfxsize=10cm
\epsfysize=10cm
\centerline{\epsffile{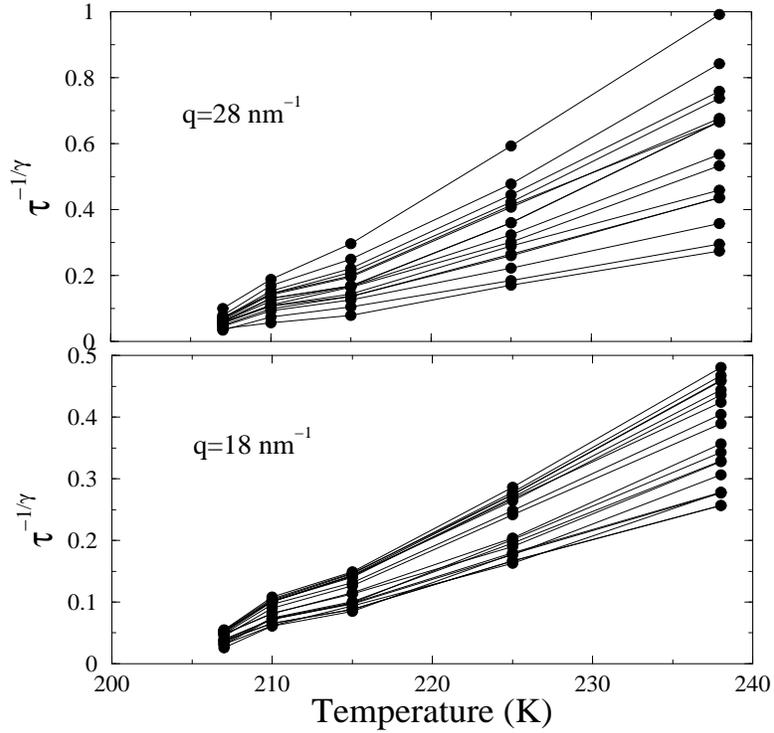}}
\caption{ $\alpha$-relaxation time to the power ${-1/\gamma}$ 
for all correlation functions
with $l \le2$, $ -l \le m \le l$ and $ -l \le n \le l$ 
 at two different $q$-vector values. Lines are drawn to guide the eyes.
}
\label{fig:tau}
\end{figure}

In order to verify the validity of the scaling law for the 
time scale $\tau(T)$
and the ``universality'' of the exponent $\gamma$ we represent in 
Fig. \ref{fig:tau} $\tau^{-1/\gamma}$, for $\gamma=2.7$,
as a function of $T$. The two figures are at two different values of the
$q$ vector, i.e. $q=18nm^{-1}$ and $q=28nm^{-1}$ and in each figure 
$\tau^{-1/\gamma}(T)$ is shown for all examined correlation functions 
(i.e. $l \le 2$, $-l \le m \le l$  and $ -l \le n \le l $).
Data in Fig. \ref{fig:tau} suggest that,
as previously observed for the center of mass and $q$-independent rotational 
correlators, the power law (\ref{eq:gamma}) is well satisfied with the 
same ``universal'' values of the critical 
exponent $\gamma=2.7$ and of the 
critical temperature $T_c=200 \pm 3$. We also note that the time
scales for fixed $T$ vary by about one decade as can be observed by
the rather different slopes.
This analysis strongly confirms the MCT and MMCT prediction
of the existence of a unique critical temperature $T_c$ 
at which both translational {\em and} rotational degrees of 
freedom cross from an ergodic to a nonergodic dynamics following
a power law behavior ruled by an ``universal'' exponent $\gamma$. 
Thus, MMCT seems to be a good framework 
in order to describe qualitatively
and semi-quantitatively the temperature dependence of the structural 
correlations in a supercooled molecular 
liquid in a wide range of temperatures above 
a critical $T_c$, as defined by eq.(\ref{eq:gamma}).

Data in Fig.~\ref{fig:ttspUZZUZZ1}, \ref{fig:ttspUZZUZZ2} 
and \ref{fig:ttspUUZUUZ1} show oscillations at short time 
connected with the
librational and vibrational dynamics of the water molecules. 
These motions
(with a characteristic timescale of the order of 0.1 ps) modulate
the approach to the plateau and may interfere with the  universal
dynamics characteristic of the $\beta$-region \cite{osc}.
\onecolumn
\begin{multicols}{2}
If the temperature is lowered very close to $T_c$, 
MCT provides quantitative
predictions for the time evolution of the correlation function. We
stress again that these predictions are the same for MMCT.
In the region of the first scaling law MMCT predicts for
$S^m_{ln,l'n'}(q,t)$ the factorization of the time--dependence from
the dependence on $q,l,n,l',n'$ and $m$, i.e. near to $T_c$ its
time--dependence is given by the so--called $\beta$--correlator $G(t)$
which describes the dynamics close to the plateau on a time scale
$t_\sigma(T)$ ($\beta$--regime). The equation for $G(t)$ does not
depend on $q,l,n,l',n'$ and $m$, it only involves the exponent
parameter $\lambda$. This equation which is the same for MCT and MMCT
can be solved exactly in the asymptotic limits $t \ll t_\sigma$ and $t
\gg t_\sigma$ yielding for both cases a power law dependence with
exponent $a$ (critical law) and exponent $b$ (von Schweidler law),
respectively. The first one describes the relaxation onto and the
second one from the plateau.

%In the time region where the correlation 
%function remains close to the plateau
%($\beta$-region) the MCT equations can be exactly solved. The analytic
%solution, usually referred to as 
%$\beta$-correlator, can be shown to depend
%on a single parameter $\lambda$. The $\beta$-correlator is 
%essentially a matching
%of two power laws in $t$, one governing the approach to the 
%plateau (defined by an exponent named $a$) and one leading the 
%after-plateau behavior (exponent $b$). 
The two exponents are both functions of $\lambda$ according to
\cite{gotze85}

\begin{equation}
\lambda = {{\Gamma(1-a)^2} \over {\Gamma(1-2a)}}=
{{\Gamma(1+b)^2} \over {\Gamma(1+2b)}}
\label{eq:lambdaab}
\end{equation}
\noindent
where $\Gamma$ is the Euler gamma function, and they are connected to
the relaxation time exponent $\gamma$ by the relation
\begin{equation}
\gamma ={1 \over 2a}+{1 \over 2b}
\label{eq:abgamma}
\end{equation}
MMCT provides an explicit expression for $\lambda$ which contains the
static molecular correlators at $T_c$ \cite{winkler}.

The von Schweidler law which also describes the early
$\alpha$--relaxation regime is given by
\begin{eqnarray}
S^m_{ln,l'n'}(q,t) - F^{m,c}_{ln,l'n'}(q) &\cong& -H^{(1)
m}_{ln,l'n'}(q) \hat{t}^b + H^{(2) m}_{ln,l'n'}(q) \hat{t}^{2b} 
\nonumber \\
& & + O(\hat{t}^{3b})
\label{eq:vs}
\end{eqnarray}
where $F^{m,c}_{ln,l'n'}(q), \; H^{(1) m}_{ln,l'n'}(q)$ and $H^{(2)
m}_{ln,l'n'}(q)$ are, respectively, the critical nonergodicity
parameter, the critical amplitude and the amplitude of the next to
leading order correction for the unnormalized correlators. An
explicit expression for $h^{(2)}(q) = H^{(2)}(q)/S(q)$ for simple
liquids has been recently derived \cite{mayr}. The time window in
which the asymptotic
%
%Also, MCT describes the behavior in the early $\alpha$ region 
%as a power law (von Schweidler law) plus corrections:
% 
%\begin{equation}
%\phi(\hat{t}) \sim f-h^{(1)} 
%\hat{t}^b+h^{(2)} \hat{t}^{2b}+ O (\hat{t}^{3b})
%\label{eq:vs}
%\end{equation}   
%
%\noindent
%where $\hat{t}$ is $t$ 
%over a characteristic $\alpha$-relaxation time scale.
%According to MCT predictions the time window where the 
power law $\hat{t}^b$ holds is 
strongly correlator and $q$ dependent. Hence to
perform a careful MCT analysis it is always necessary to take into
account also the second 
order corrections. This has previously been found
for the center of mass correlator for water \cite{collective} and for the
molecular correlators for diatomic molecules \cite{kks}.
In order to test the validity of the von Schweidler law and 
the relevance of the second order corrections we have fitted the
time evolution of our correlators according to eq.(\ref{eq:vs})
for the lowest simulated temperature 
$T=207 K$ which is few degrees above the critical one.
\end{multicols}
\begin{figure}
\epsfxsize=10cm
\epsfysize=8cm
\centerline{\epsffile{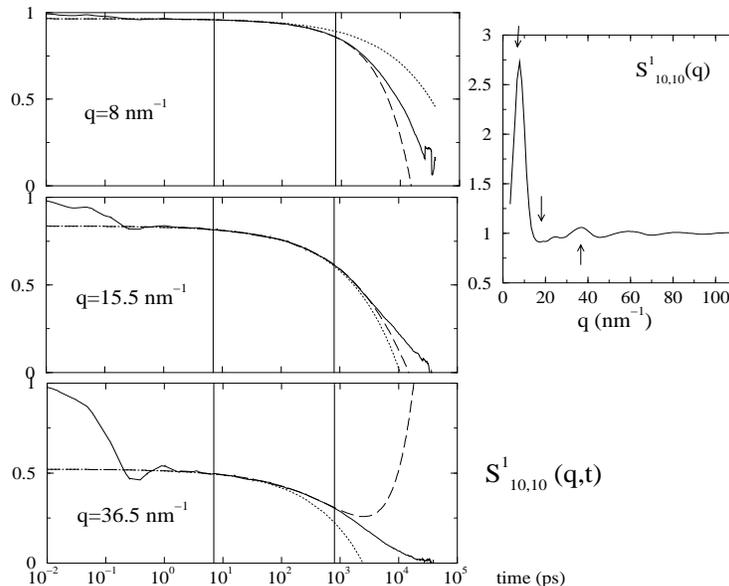}}
\caption{ Fit to eq.(\protect\ref{eq:vs}) of the time dependence of 
 $S^m_{ln,l'n'}(q,t)/S^m_{ln,l'n'}(q)$ for different $q$-vector values.
 The solid line represents the MD--data whereas the dashed line and
 the dotted line are fits to the von Schweidler law with and without
 second order correction, respectively.
 The inset shows the corresponding static structure factor with an 
 indication of the chosen $q$-vectors (see arrows). 
 This figure refers to  $S^1_{10,10}(q,t)$
}
\label{fig:fitUUZUUZ}
\end{figure}
\begin{figure}
\epsfxsize=10cm
\epsfysize=8cm
\centerline{\epsffile{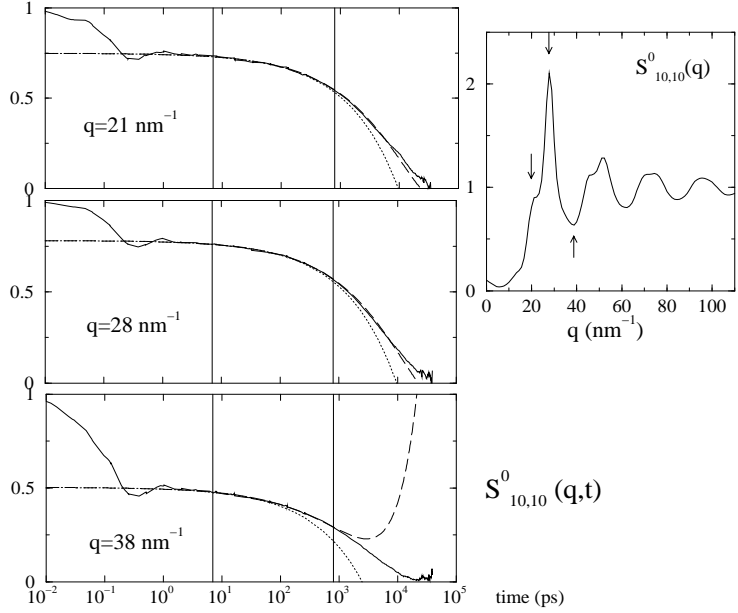}}
\caption{ Same as \protect\ref{fig:fitUUZUUZ} for $S^0_{10,10}(q,t)$
}
\label{fig:fitUZZUZZ}
\end{figure}
\begin{multicols}{2}
In Figs. \ref{fig:fitUUZUUZ}-
%, \ref{fig:fitUZZUZZ}, \ref{fig:fitDZZDZZ}, 
\ref{fig:fitDZZZZZ} we show the
fits with eq.(\ref{eq:vs}) performed for some representative normalized
diagonal
correlators. The vertical lines indicate the time window selected for
the fit which has been chosen consistently with the previous fits for
translational and $q$-independent correlators, i.e $t$ covers a range
of two orders of magnitude, from $7 ps$ up to $800 ps$.
Fig. \ref{fig:fitUUZUUZ} represents the fit with eq.(\ref{eq:vs})
using the value $b=0.5$ as previously found in 
the MCT study of the time dependence of self and 
collective center of mass correlators. We show the
correlator $S^1_{10,10}(q,t)$ for three different values of the $q$
vector, marked by arrows in the small graph representing the static
$S^1_{10,10}(q)$. The solid line is the numerical curve, while the
long dashed line is the result of the fit. The quality of the fit is
remarkably good. The $q$-dependence of the validity of the fit is
evident noting that for $q=8nm^{-1}$ and $q=15.5 nm^{-1}$ the fitted von
Schweidler law holds far above the fitting range, 
while for $q=36.5 nm^{-1}$
the two curves separate out of the fit window.  
\end{multicols}
\begin{figure}
\epsfxsize=10cm
\epsfysize=8cm
\centerline{\epsffile{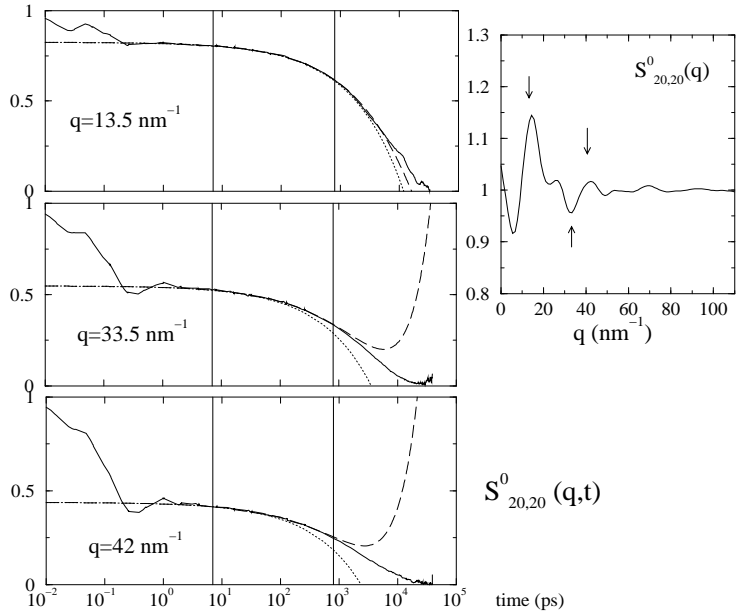}}
\caption{Same as \protect\ref{fig:fitUUZUUZ} for $S^0_{20,20}(q,t)$
}
\label{fig:fitDZZDZZ}
\end{figure}
\begin{figure}
\epsfxsize=10cm
\epsfysize=8cm
\centerline{\epsffile{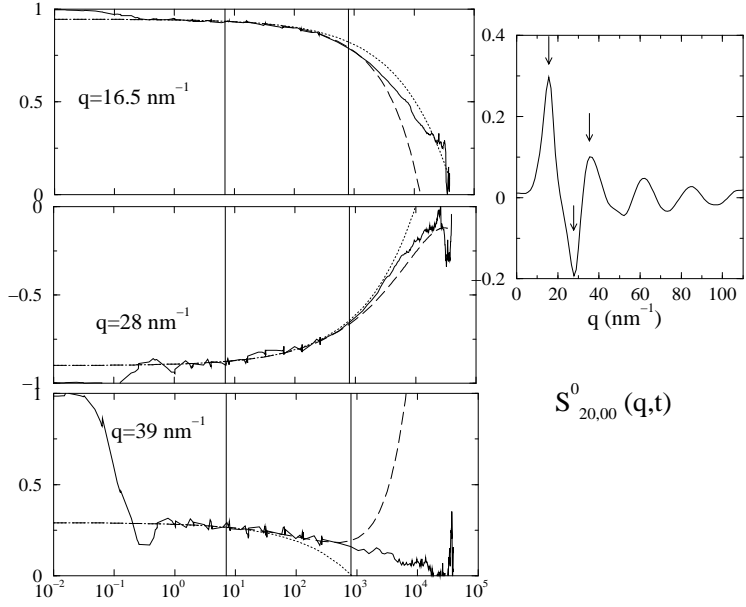}}
\caption{Same as \protect\ref{fig:fitUUZUUZ} for $S^0_{20,00}(q,t)$
}
\label{fig:fitDZZZZZ}
\end{figure}
\begin{multicols}{2}
To clarify the
importance of the second order corrections we have also reported in
the figure the curve obtained keeping only the term $\hat{t}^b$ (short
dashed line). We see that while in the case 
$q=15.5 nm^{-1}$ the first power
law fit alone extends to almost three order of magnitude, for the
other values of $q$ the second order corrections are necessary.
The obtained fitting parameters 
$F$, $H^{(1)}$ and $H^{(2)}$ are shown in
Fig. \ref{fig:par-vs} as functions of $q$ together 
with the static structure factors.
Analytic expressions for the 
calculation of these quantities are provided by
MMCT (for the amplitude $H^{(2)}$ no explicit expression for the
molecular system has been derived yet), so that they can be, 
in principle, calculated using the molecular 
static structure factors. For diatomic molecules this has been
done for $F$ and $H^{(1)}$ \cite{winkler}.

The fit to other correlators show similar behaviors. 
We want to stress that, according to MCT and MMCT predictions, 
in all the fits to eq.(\ref{eq:vs}) the same exponent $b$ has 
turned out to be satisfactory. The independence of the power 
law exponent $b$ on the chosen correlator and on the 
$q$ value  is a strong argument in favor of MMCT as
framework to describe the slow relaxation in
supercooled molecular liquids. We also stress that the 
two independently calculated exponents $b$ and $\gamma$ satisfy the
theoretical prediction of eq.(\ref{eq:abgamma}).

In view of a future comparison with the full time dependence of the 
MMCT correlations, we report in 
Fig.~\ref{fig:par-kww} the parameters of the
fit to the numerical correlation function according 
to a stretched exponential form 
\begin{eqnarray}
S^m_{ln,l'n'}(q,t) &=& A^m_{ln,l'n'}(q) \; \cdot \nonumber \\
& & \; \cdot \;\; \exp\left[ 
-\left(\frac{t}{\tau^m_{ln,l'n'}(q,T)} \right)^{\beta^m_{ln,l'n'}(q)}
\right].
\label{eq:kww}
\end{eqnarray}

We note that
in all examined cases the large $q$-vector limit of the 
stretching parameter $\beta$ is about $0.5$, i.e. equal to 
the value of $b$. Such equivalence is 
predicted by MCT\cite{fuchs-beta}.
\end{multicols}
\begin{figure}
\centerline{
\epsfxsize=7cm
\epsfysize=7cm
\epsffile{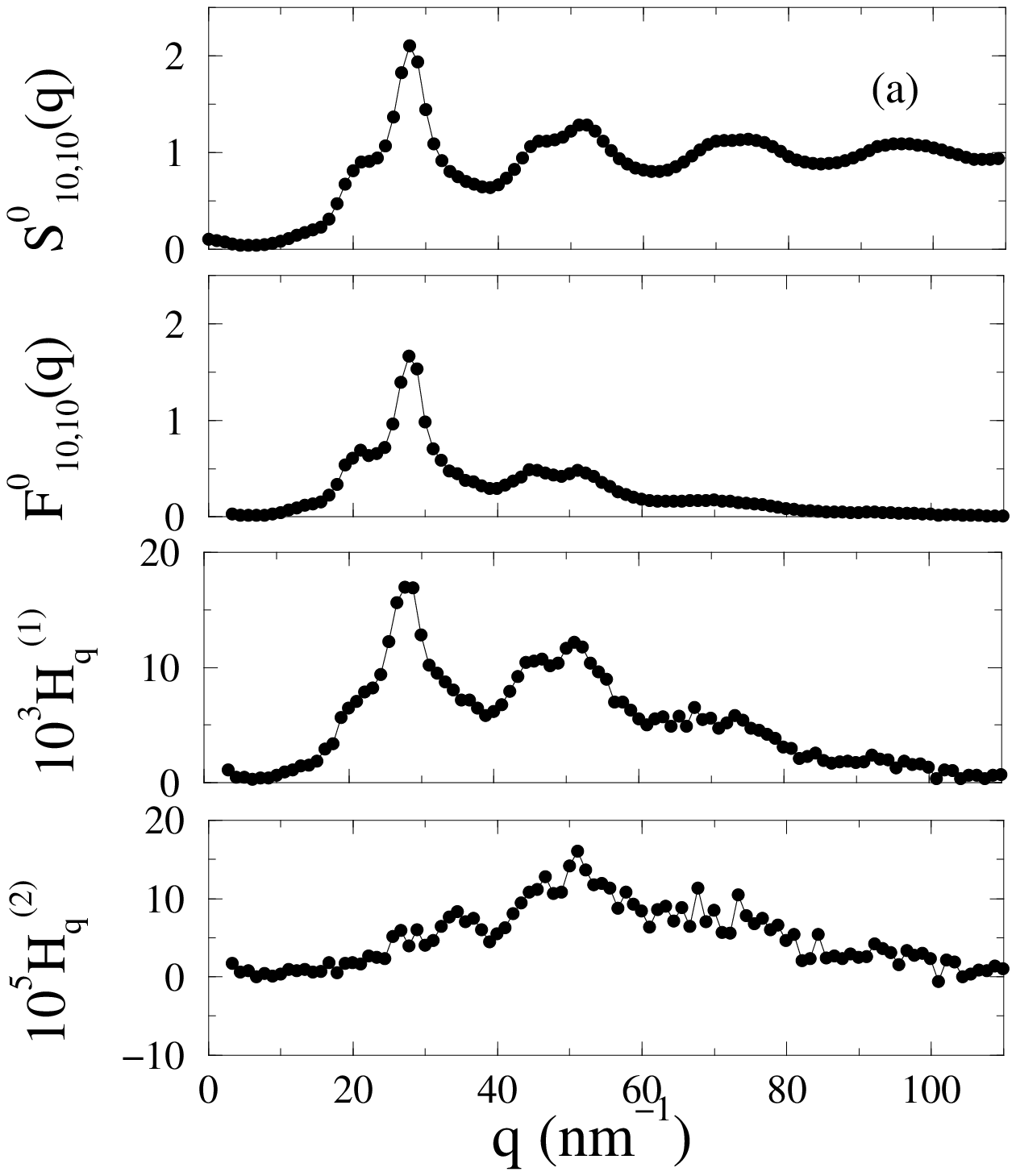}
\epsfxsize=7cm
\epsfysize=7cm
\epsffile{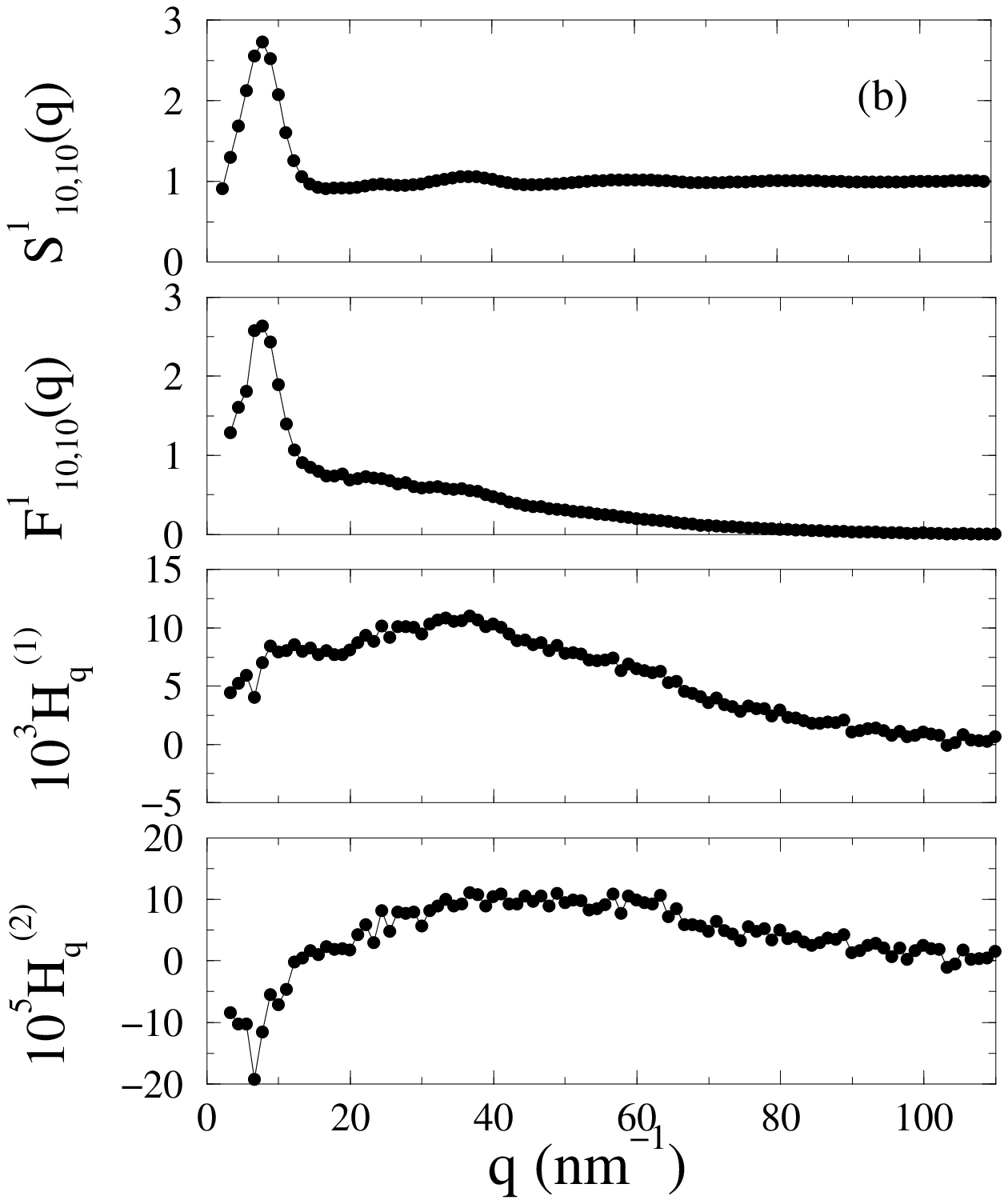}
}
\centerline{
\epsfxsize=7cm
\epsfysize=7cm
\epsffile{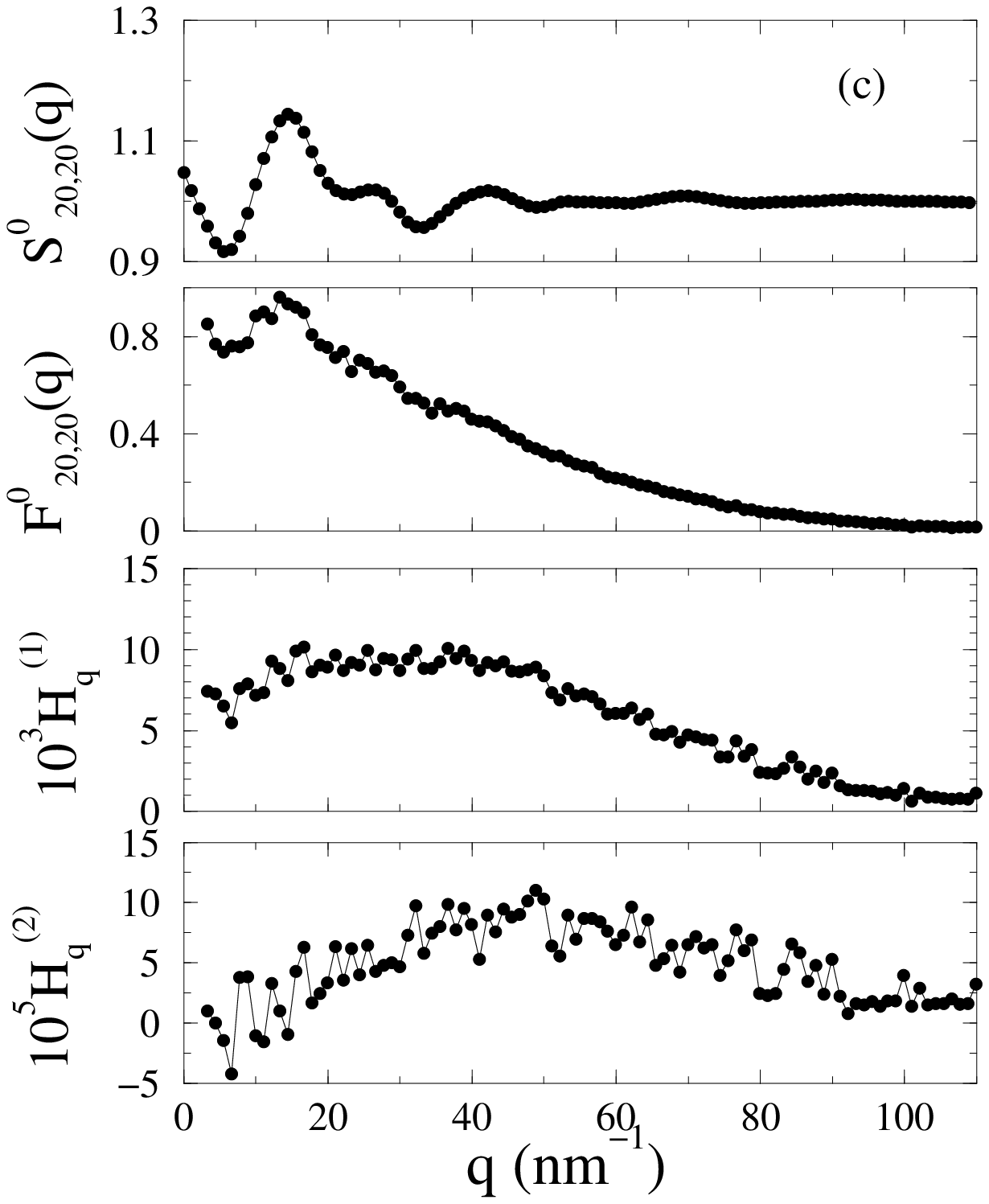}
\epsfxsize=7cm
\epsfysize=7cm
\epsffile{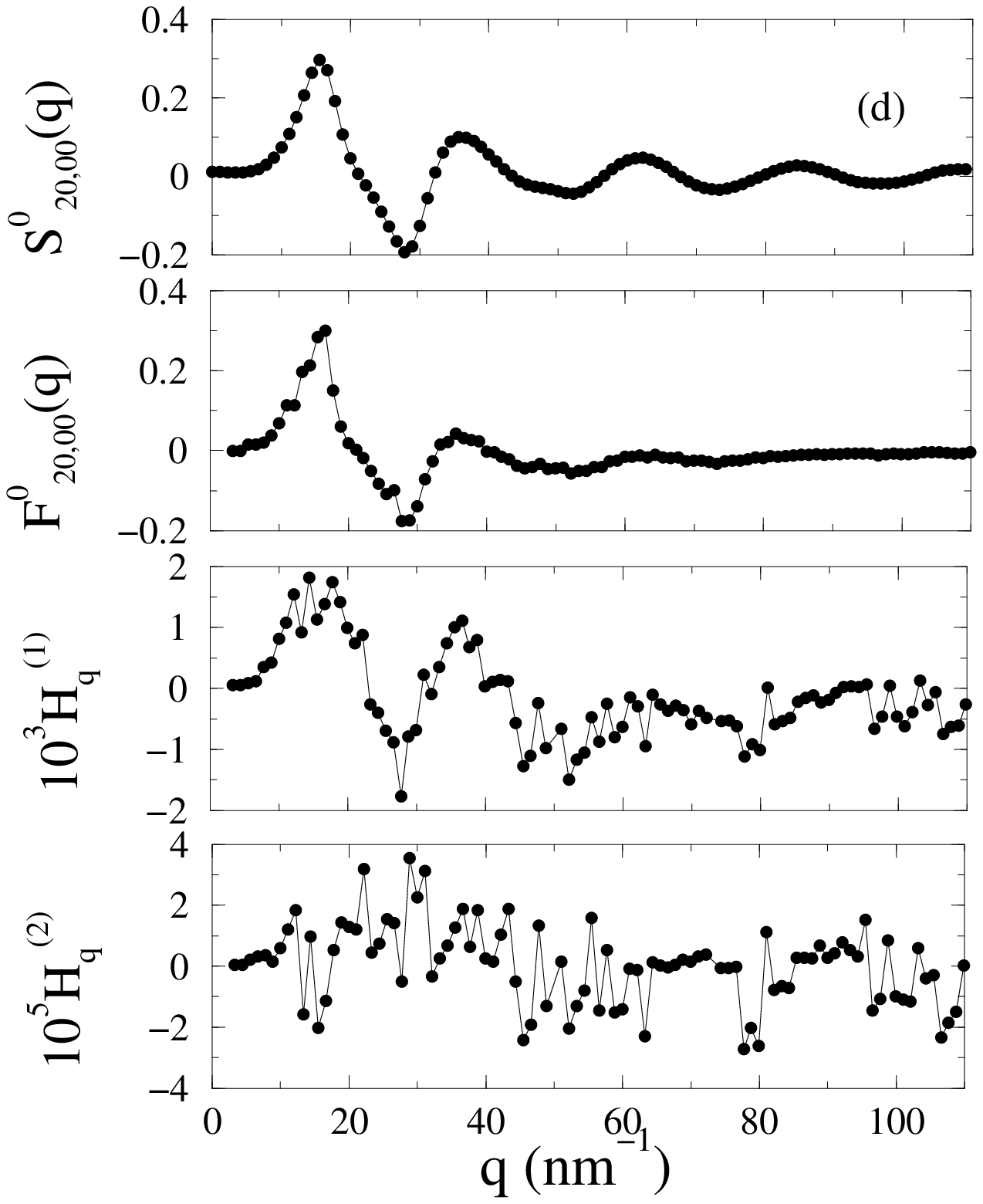}
}
\caption{Parameters of the fits 
shown in Figs.\protect\ref{fig:fitUUZUUZ}-
\protect\ref{fig:fitDZZZZZ}.  The parameters refer to fit of the
unnormalized  $S^m_{ln,l'n'}(q,t)$. 
Note that the statistical noise is large
when the amplitude of the $\alpha$ relaxation is rather small.
}
\label{fig:par-vs}
\end{figure}
\newpage
\begin{figure}
\centerline{
\epsfxsize=7cm
\epsfysize=7cm
\epsffile{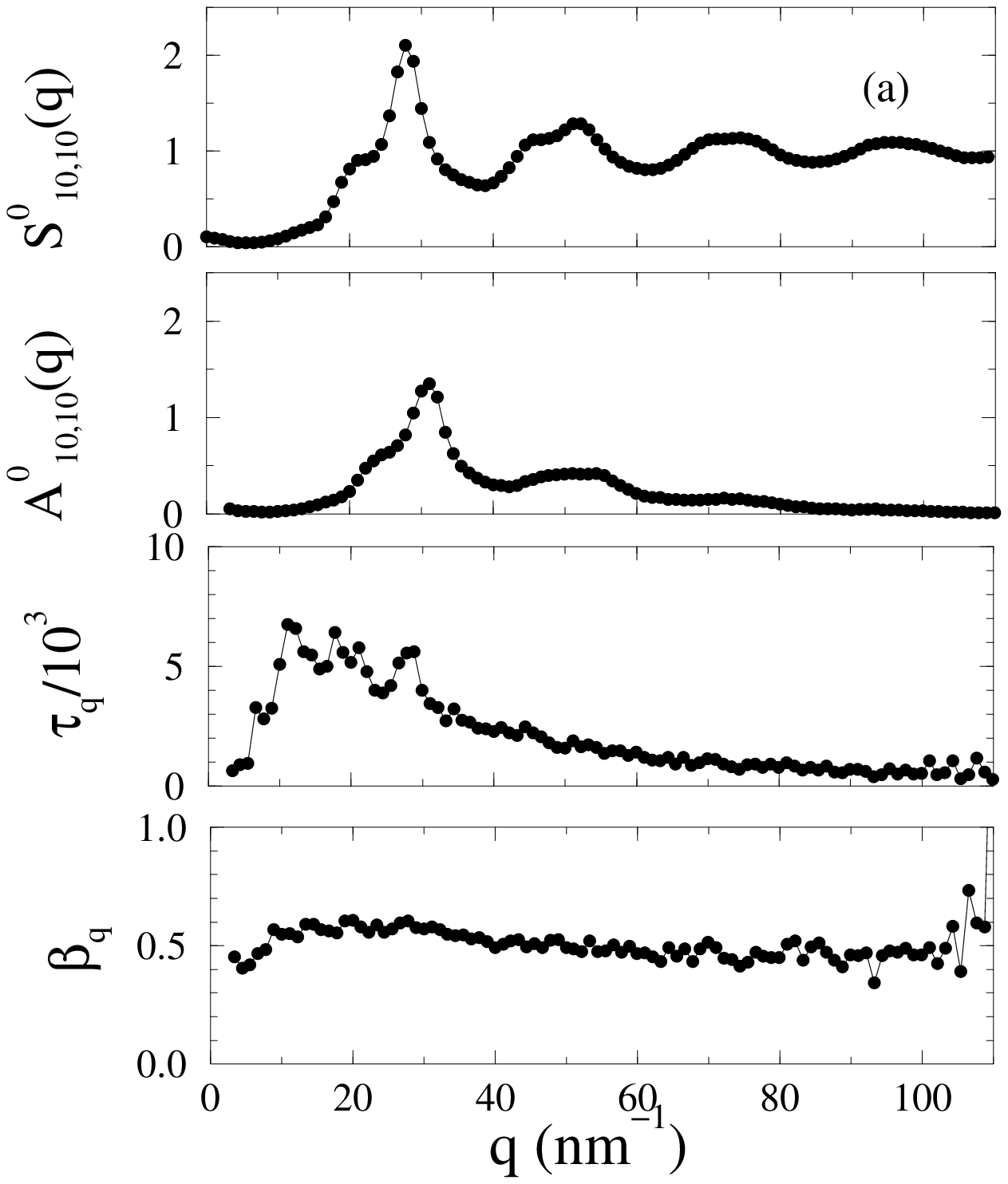}
\epsfxsize=7cm
\epsfysize=7cm
\epsffile{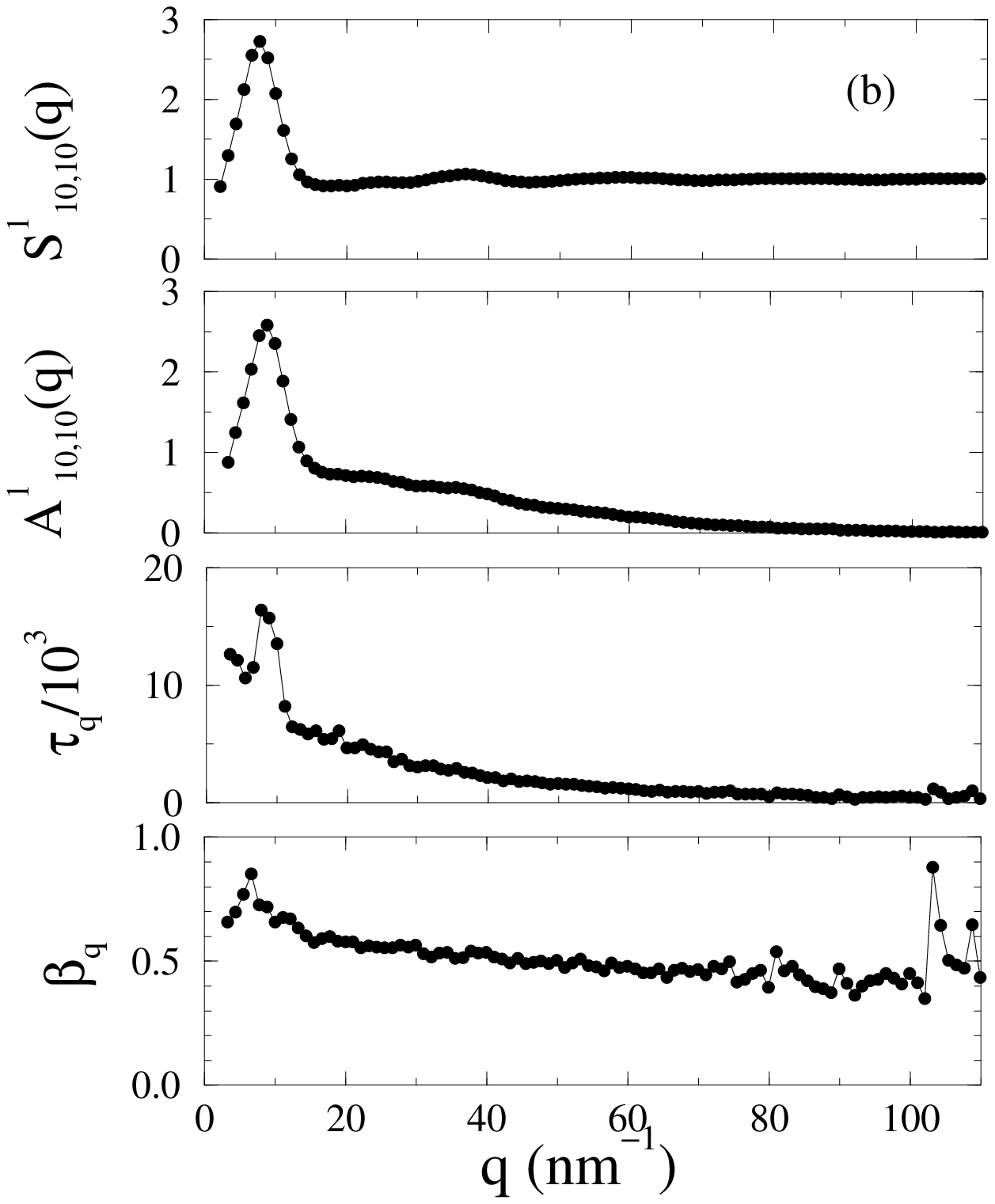}
}
\centerline{
\epsfxsize=7cm
\epsfysize=7cm
\epsffile{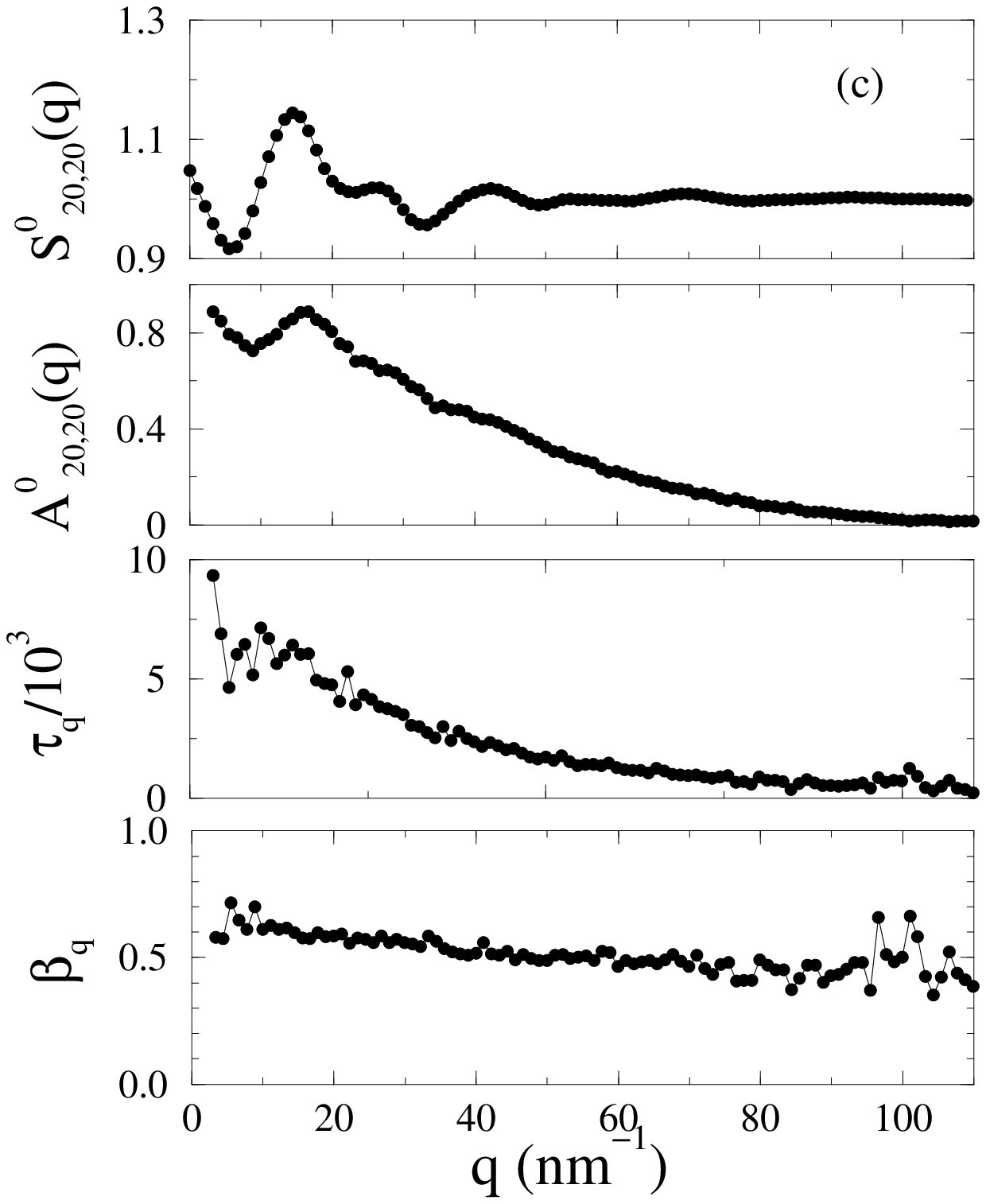}
\epsfxsize=7cm
\epsfysize=7cm
\epsffile{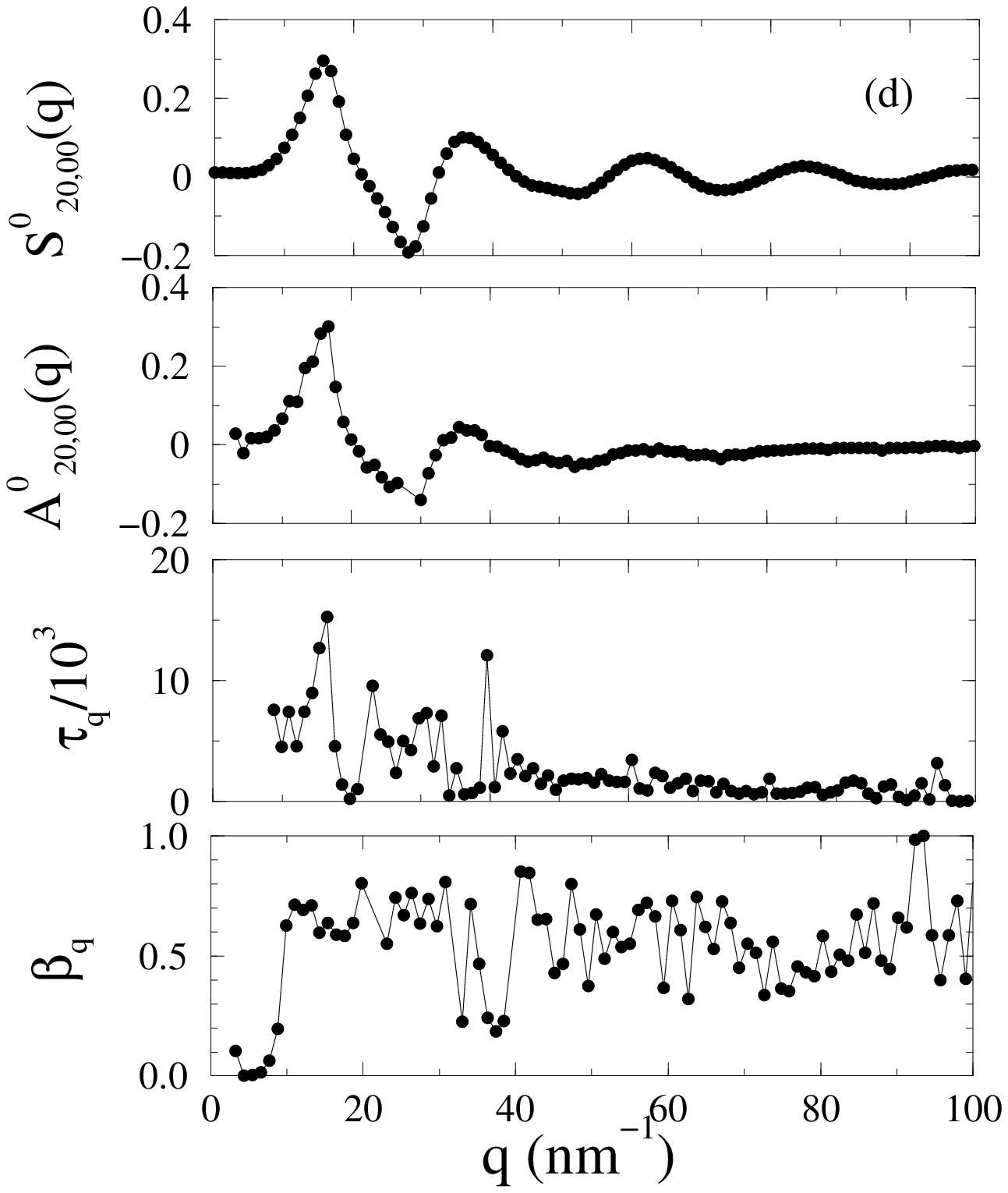}
}
\caption{ Parameters of the fits to a 
stretched exponential function of the
same correlators shown in
Figs.\protect\ref{fig:fitUUZUUZ}-
\protect\ref{fig:fitDZZZZZ}.  The parameters refer to fit of the
unnormalized  
$S^m_{ln,l'n'}(q,t)$. Note that the statistical noise is large
when the amplitude of the $\alpha$ relaxation is rather small.
}
\label{fig:par-kww}
\end{figure}
\twocolumn

\section{summary and conclusions}

In this paper we have analyzed 
the slow relaxation properties of a supercooled
liquid of planar rigid molecules described by means of a MD
simulation. The parameters of the simulation have been chosen in such
a way as to mimic the thermodynamic and dynamic properties of liquid
water, a molecular liquid characterized by a network structure at low
temperatures.
Starting from the set of MD data we have calculated all the static and
dynamic molecular structure factors $S^m_{ln,l'n'}(q,t)$ up to $l=2$
in the $q$-frame. These correlators present a variety of features which 
open a deeper insight into the structural properties of the liquid.  
For example the symmetries of the rotational correlators reflect the 
geometrical properties of the 
water molecules, while the peak distribution
describes the intermolecular 
interactions. It becomes clear from an overview
on the complete set of correlators that a theory able to describe the
relaxational properties of a 
supercooled molecular liquid can not neglect
the molecular correlators with $l,l' \not= 0$ and 
their coupling to the center of mass one. For the static distinct part at 
least at high
temperatures we have proven that it is independent of the sign of $n$
and $n'$ for molecules with $C_{rv}$ symmetry. The MD--results
indicate that the independence even holds at lower temperatures (down
to $207 K$) within the numerical error.

In this Article we argue that MCT in its molecular formulation is 
a good candidate to be the correct framework for the description
of the slow dynamics in supercooled molecular liquids. 
In order to support this
view we have compared the properties of our MD liquid to the 
universal predictions of MCT (or MMCT) which are independent of the
molecular nature of the liquid. 
Even if the complete set of MMCT equations has 
not yet been solved, thus preventing from a rigorous comparison between
theory and simulation, we have tested several 
asymptotic MCT predictions.
We have shown that
the temperature and time dependence of our MD correlators 
perfectly obeys the time-temperature superposition principle predicted by MCT and MMCT, and 
we have found the MMCT ``universality'', i.e. the critical temperature
$T_c$ the exponents $b$ and $\gamma$ (and therefore also $a$) do not
depend on $q,l,n,l',n'$ and $m$.

Close to $T_c$, where the asymptotic $\beta$-correlator behavior 
is reinforced, we have compared the
numerical correlator with the predicted von Schweidler law plus second
order corrections in the early 
$\alpha$-region. We have shown good quality 
fits consistent with what was 
previously found for the self and collective 
dynamics of the center of mass alone and 
of the $q$-independent angular correlators. 
Again, the predicted relation (\ref{eq:abgamma}) between the 
von Schweidler exponent and 
the relaxation time exponent is  fulfilled, 
as well as the large $q$-vector
limit of all stretching exponent $\beta^m_{ln,l'n'}(q)$.

\acknowledgments
The work of L. Fabbian, F. Sciortino and P. Tartaglia 
has been partially supported by  MURST (PRIN 97). A.Latz,
R.Schilling and C.Theis are grateful for financial support by SFB--262.

\begin{appendix}
\section{Molecules with $C_{rv}$--symmetry}
\label{appsym}

In this appendix we will investigate the validity of the $n
\to -n$ symmetry for the {\em static} distinct part of the
correlators (c.f. eq.(\ref{a21})) for molecules with
$C_{rv}$--symmetry.

We consider a liquid of $N$ identical, rigid molecules. Each molecule
is made up of $N_\alpha$ atoms of type $\alpha$, $\alpha = 1,2,...$.
The position of atom $\nu_\alpha$ of type $\alpha$ in the $j$th
molecule is denoted by $\vec{x}_{j,\nu_\alpha}^{(\alpha)}$, $j =
1,2,...,N$, $\nu_\alpha = 1,2,...,N_\alpha$. Then, in a site--site
representation the total potential energy $V$ is given by:
\begin{equation}
V(\{\vec{x}_{j,\nu_\alpha}^{(\alpha)}\}) = \frac{1}{2} \sum_{j \not=j'}
\sum_{\alpha,\alpha'} \sum_{\nu_\alpha, \nu'_{\alpha'}} v_{\alpha
\alpha'}(|\vec{x}_{j,\nu_\alpha}^{(\alpha)}-
\vec{x}_{j',\nu'_{\alpha'}}^{(\alpha')}|)
\label{eq:app1}
\end{equation}
where we restrict ourselves to two--body interactions with pair
potential $v_{\alpha \alpha'}(\vec{x})$ between atoms of type $\alpha$
and $\alpha'$. Introducing the microscopic molecular density for atoms
of type $\alpha$ in molecule $j$:
\begin{equation}
\rho_j^{(\alpha)}(\vec{x}) = \sum_{\nu_\alpha=1}^{N_\alpha}
\delta(\vec{x} - \vec{x}_{j,\nu_\alpha}^{(\alpha)})
\label{eq:app2}
\end{equation}
and its Fourier transform:
\begin{equation}
\rho_j^{(\alpha)}(\vec{q}) = \sum_{\nu_\alpha=1}^{N_\alpha}
e^{i \vec{q} \cdot \vec{x}_{j,\nu_\alpha}^{(\alpha)}}
\label{eq:app3}
\end{equation}
we can rewrite (\ref{eq:app1}) as follows:
\begin{eqnarray}
& &V(\{\vec{x}_{j,\nu_\alpha}^{(\alpha)}\}) = 
\frac{1}{2} \sum_{j \not=j'}
\sum_{\alpha,\alpha'} \int_V d^3x \int_V d^3x'  \; \cdot \nonumber \\
& & \quad \cdot \; \rho_j^{(\alpha)}(\vec{x}) 
v_{\alpha \alpha'}(\vec{x}-\vec{x'})
\rho_{j'}^{(\alpha')}(\vec{x'})
\label{eq:app4}
\end{eqnarray}
or by use of (\ref{eq:app3}):
\begin{eqnarray}
& &V(\{\vec{x}_{j,\nu_\alpha}^{(\alpha)}\}) = 
\frac{1}{2} \sum_{j \not=j'}
\sum_{\alpha,\alpha'} \frac{1}{V} \sum_{\vec{q}} \; \cdot \nonumber \\
& & \quad \cdot \; \rho_j^{(\alpha)
\ast}(\vec{q}) \tilde{v}_{\alpha \alpha'}(\vec{q})
\rho_{j'}^{(\alpha')}(\vec{q})
\label{eq:app5}
\end{eqnarray}
with $\tilde{v}_{\alpha \alpha'}(\vec{q}) = \int d^3x \; v_{\alpha
\alpha'}(\vec{x}) e^{i \vec{q} \cdot \vec{x}}$, the Fourier transform
of the pair potential which depends on $q = |\vec{q}|$ and the
volume $V$ of the system only.

Instead of the site--site coordinates we introduce center of mass and
relative coordinates:
\begin{equation}
\vec{x}_j = \left( \sum_\alpha m_\alpha \sum_{\nu_\alpha=1}^{N_\alpha}
\vec{x}_{j,\nu_\alpha}^{(\alpha)} \right) / 
\sum_\alpha N_\alpha m_\alpha
\label{eq:app6}
\end{equation}
and
\begin{equation}
\vec{r}_{j,\nu_\alpha}^{(\alpha)} = \vec{x}_{j,\nu_\alpha}^{(\alpha)} -
\vec{x}_j ,
\label{eq:app7}
\end{equation}
respectively. $m_\alpha$ is the mass of atoms of type $\alpha$. Next
we substitute (\ref{eq:app6}) 
and (\ref{eq:app7}) into (\ref{eq:app3}) which leads to
\begin{equation}
\rho_j^{(\alpha)}(\vec{q}) = e^{i \vec{q} \cdot \vec{x}_j}
\tilde{\rho}_j^{(\alpha)}(\vec{q})
\label{eq:app8}
\end{equation}
where
\begin{equation}
\tilde{\rho}_j^{(\alpha)}(\vec{q}) = \sum_{\nu_\alpha=1}^{N_\alpha}
e^{i \vec{q} \cdot \vec{r}_{j,\nu_\alpha}^{(\alpha)}} .
\label{eq:app9}
\end{equation}
Now, let $\{\vec{R}_{\nu_\alpha}^{(\alpha)}\}$ be the relative vectors
in the body fixed frame and $\Omega_j = (\phi_j, \theta_j, \chi_j)$
the Euler angles between the laboratory and the body fixed frame. Then
it is:
\begin{equation}
\vec{r}_{j,\nu_\alpha}^{(\alpha)} = R^{-1}(\Omega_j)
\vec{R}_{\nu_\alpha}^{(\alpha)}
\label{eq:app10}
\end{equation}
with $R(\Omega) \in SO(3)$. Substitution of (\ref{eq:app10}) into
(\ref{eq:app9}) and making use of the Rayleigh expansion \cite{G&G} we
arrive at
\begin{eqnarray}
& &\tilde{\rho}_j^{(\alpha)}(\vec{q},\Omega_j) = 4 \pi \sum_{lm} i^l
j_l(q R_\alpha) Y_{lm}^\ast(R(\Omega_j) \vec{e}_q) \; \cdot \nonumber \\
& & \quad \cdot \; \sum_{\nu_\alpha=1}^{N_\alpha} 
Y_{lm}(\vec{e}_{\nu_\alpha}^{(\alpha)})
\label{eq:app11}
\end{eqnarray}
and with the transformation of the spherical harmonics under $R \in
SO(3)$ \cite{G&G}:
\begin{eqnarray}
& &\tilde{\rho}_j^{(\alpha)}(\vec{q},\Omega_j) = 4 \pi \sum_{lm} i^l
j_l(q R_\alpha) D_{m'm}^{l \ast}(\Omega_j) Y_{lm}^\ast(\vec{e}_q) 
\; \cdot \nonumber \\
& & \quad \cdot \; \sum_{\nu_\alpha=1}^{N_\alpha}
Y_{lm}(\vec{e}_{\nu_\alpha}^{(\alpha)}).
\label{eq:app12}
\end{eqnarray}
$j_l(r)$ are the spherical Bessel functions, $D_{m'm}^{l
}(\Omega)$ are Wigner's rotation matrices, $R_\alpha =
|\vec{R}_{\nu_\alpha}^{(\alpha)}|$ for {\em all} $\nu_\alpha$,
$\vec{e}_{\nu_\alpha}^{(\alpha)} =
\vec{R}_{\nu_\alpha}^{(\alpha)}/R_\alpha$ and $\vec{e}_q = \vec{q}/q$.
The substitution of $\rho_j^{(\alpha)}(\vec{q})$ (\ref{eq:app8}) with
$\tilde{\rho}_j^{(\alpha)}(\vec{q},\Omega_j)$ from (\ref{eq:app12})
into (\ref{eq:app5}) yields the center of mass--angular representation of $V$.
\begin{eqnarray}
V(\{\vec{x}_{j, \nu_\alpha}^{(\alpha)}\}) &\equiv&
V(\vec{x}_1,...,\vec{x}_N;\Omega_1,...,\Omega_N) = \nonumber \\
& & \frac{1}{2} \sum_{j \not= j'} \frac{1}{V} \sum_{\vec{q}}
\sum_{lmn \atop l'm'n'} \tilde{v}_{lmn,l'm'n'}(\vec{q}) 
\; \cdot \nonumber \\ 
& & \cdot \; e^{-i \vec{q} \cdot (\vec{x}_j - \vec{x}_{j'})} D_{mn}^l(\Omega_j)
D_{m'n'}^{l' \ast}(\Omega_{j'})
\label{eq:app13}
\end{eqnarray}
with the transformed pair potential
\begin{eqnarray}
& &\tilde{v}_{lmn,l'm'n'}(\vec{q}) = (4\pi)^2 i^{l'-l}
Y_{lm}(\vec{e}_q) Y^\ast_{l'm'}(\vec{e}_q) \sum_{\alpha \alpha'}
\; cdot \nonumber \\
& & \; \cdot \quad \tilde{v}_{\alpha \alpha'}(q) j_l(q R_\alpha) 
j_{l'}(q R_{\alpha'}) \; \cdot \nonumber \\ 
& & \; \cdot \; \left( \sum_{\nu_\alpha=1}^{N_\alpha}
Y_{ln}(\vec{e}_{\nu_\alpha}^{(\alpha)}) \right) \left(
\sum_{\nu'_{\alpha'}=1}^{N_\alpha} 
Y_{l'n'}(\vec{e}_{\nu'_{\alpha'}}^{(\alpha')})
\right).
\label{eq:app14}
\end{eqnarray}
The representation of $V$ by (\ref{eq:app13}) is quite obvious, since
the interaction potential $V_0(\vec{x},\Omega,\Omega')$ between 
{\em two}
molecules with center of mass separation $\vec{x}$ and orientation $\Omega$ and
$\Omega'$ can be expanded with respect to the complete set of
functions $e^{i\vec{q}\cdot\vec{x}} D_{mn}^l(\Omega) D_{m'n'}^{l'
\ast}(\Omega')$.

All the local molecular symmetry is contained in the coefficients
$\tilde{v}_{lmn,l'm'n'}(\vec{q})$ through the entity:
\begin{equation}
y_{ln}^{(\alpha)} = \sum_{\nu_\alpha=1}^{N_\alpha}
Y_{ln}(\vec{e}_{\nu_\alpha}^{(\alpha)}),
\label{eq:app15}
\end{equation}
which will be discussed for $C_{rv}$ symmetry. In the following we
choose the body fixed $z$--axis along the molecular $r$--fold symmetry
axes. First of all we note that the {\em $r$--fold rotational
symmetry} implies:
\begin{equation}
y_{ln}^{(\alpha)} = \left\{
\begin{array}{c@{\;\; , \;}c}
\not= 0 & n \in \{0,\pm r, \pm 2r, ...\} \\
= 0 & \mbox{otherwise} 
\end{array}
\right.
\label{eq:app16}
\end{equation}
Here we have assumed that {\em all} types of atoms $\alpha = 1,2,...$
have exactly an $r$--fold rotational symmetry. This is not necessary
for a molecule with $r$--fold axes. There may be some types of atoms
which have a $2r$--, $3r$--, etc. fold symmetry. For such $\alpha$ it
is $y_{ln}^{(\alpha)} \not= 0$ for $n = \mu^{(\alpha)} r$ where
$\mu^{(\alpha)} = 2,3,...$. Nevertheless, (\ref{eq:app16}) remains
true since $\mu = \min_\alpha \mu^{(\alpha)} = 1$, by assumption.

Now let us turn to the {\em reflection symmetry}. With 
$(\phi_{\nu_\alpha}^{(\alpha)}, \theta^{(\alpha)})$ (where
$\theta^{(\alpha)}$ does {\em not} depend on $\nu_\alpha$, due to the
$C_{rv}$ symmetry and the appropriate choice of the body fixed
$z$--axes) we denote the azimuthal and polar angles of
$\vec{e}_{\nu_\alpha}^{(\alpha)}$. For $r$ {\em even} and fixed
$\alpha$ the angles $\phi_{\nu_\alpha}^{(\alpha)}$ can be chosen such
that the reflection symmetry with respect to the $(x-z)$--plane
implies:
\begin{equation}
\phi_{N_\alpha-\nu_\alpha+1}^{(\alpha)} = \pi -
\phi_{\nu_\alpha}^{(\alpha)} + \lambda_{\nu_\alpha} 2 \pi
\label{eq:app17}
\end{equation}
for $\nu_\alpha = 1,2,...,N_\alpha$. The integer
$\lambda_{\nu_\alpha}$ may depend on $\nu_\alpha$. using the explicit
$(\phi,\theta)$ dependence of $Y_{ln}(\vec{e}) \equiv
Y_{ln}(\theta,\phi) = c_{ln} P_{ln}(\theta) e^{in\phi}$ \cite{G&G}, we
get from (\ref{eq:app15}) with (\ref{eq:app17}):
\begin{eqnarray}
y_{ln}^{(\alpha)} &=& c_{ln} P_{ln}(\theta^{(\alpha)})
\sum_{\nu_\alpha=1}^{N_\alpha} e^{in\phi_{\nu_\alpha}^{(\alpha)}}
\nonumber \\
&=& c_{ln} P_{ln}(\theta^{(\alpha)}) (-1)^n
\sum_{\nu_\alpha=1}^{N_\alpha}
e^{-in\phi_{N_\alpha-\nu_\alpha+1}^{(\alpha)}} \nonumber \\
&=& (-1)^n c_{ln} P_{ln}(\theta^{(\alpha)})
\sum_{\nu_\alpha=1}^{N_\alpha} e^{-in\phi_{\nu_\alpha}^{(\alpha)}}
\nonumber \\
&=& \sum_{\nu_\alpha=1}^{N_\alpha}
Y_{l\underline{n}}(\phi_{\nu_\alpha}^{(\alpha)}, \theta^{(\alpha)})
\nonumber \\
&=& y_{l\underline{n}}^{(\alpha)}
\label{eq:app18}
\end{eqnarray}
where $\underline{n} = -n$. In the case that $r$ is {\em odd} a
similar choice of $\phi_{\nu_\alpha}^{(\alpha)}$ can be done, such
that (\ref{eq:app18}) holds. Due to (\ref{eq:app18}) we find that the
coefficients $\tilde{v}_{lmn,l'm'n'}(\vec{q})$ do {\em not} depend on
the sign of $n$ and $n'$. Therefore (\ref{eq:app13}) can also be
rewritten as follows:
\begin{eqnarray}
& & V(\vec{x}_1,...,\vec{x}_N;\Omega_1,...,\Omega_N) = \nonumber \\ 
& & \quad \frac{1}{8}
\sum_{j \not= j'} \frac{1}{V} \sum_q \sum_{\kappa \kappa'} v_{\kappa
\kappa'}(\vec{q}) e^{ -i \vec{q} \cdot (\vec{x}_j - \vec{x}_{j'} ) }
\; \cdot \nonumber \\
& &\; \cdot \; [ D_\kappa (\Omega_j) + D_{\underline{\kappa}} (\Omega_j)
] [ D_{\kappa'}(\Omega_{j'}) + 
D_{\underline{\kappa}'}(\Omega_{j'}) ]^\ast
\label{eq:app19}
\end{eqnarray}
where $\kappa = (l,m,n), \; \underline{\kappa} = (l,m,\underline{n})$
and $D_\kappa = D_{mn}^l$ has been used as a shorthand notation.
(\ref{eq:app19}) is the basic result which allows to study the $n
\to -n$ symmetry for the static distinct part of the
correlators. From (\ref{eq:app6}) we get for $t = 0$:
\begin{eqnarray}
& &S_{\kappa,\kappa'}^{(d)}(\vec{q}) = i^{l'-l} \left[ (2l+1)(2l'+1)
\right]^\frac{1}{2} \frac{1}{N} \sum_{k \not= k'} \frac{1}{Z_c}
\; \cdot \nonumber \\
& & \; \cdot \; \int \prod_{j=1}^{N}(d^3x_j d\Omega_j) e^{-i \vec{q} \cdot 
(\vec{x}_k - \vec{x}_{k'})} D_\kappa(\Omega_k)
D_{\kappa'}^\ast(\Omega_{k'}) \; \cdot \nonumber \\
& &\; \cdot \; 
e^{-\beta V(\vec{x}_1,...,\vec{x}_N;\Omega_1,...,\Omega_N)} 
\label{eq:app20}
\end{eqnarray}
where $Z_c$ is the configurational partition function. Now we expand
$\exp(-\beta V)$ into a power series. The zeroth order term of
$S_{\kappa,\kappa'}^{(d)}(\vec{q})$ is proportional to $\delta_{n0}
\delta_{n'0}$ which in a trivial sense is invariant under $n$ or $n'$
into $-n$ or $-n'$, for arbitrary $\vec{q},l,m,l'$ and $m'$. For the
first order term we find with (\ref{eq:app19}):
\begin{equation}
\left( S_{\kappa,\kappa'}^{(d)}(\vec{q}) \right)_{\mbox{1.order}}
\propto \tilde{v}_{\kappa, \kappa'}(\vec{q})
\label{eq:app21}
\end{equation}
where the factor of proportionality is independent of $m,m',n$ and
$n'$. Since $\tilde{v}_{\kappa, \kappa'}$ does not depend on the sign
of $n$ and $n'$ for all $\vec{q},l,m,l'$ and $m'$, the same is true
for $S_{\kappa,\kappa'}^{(d)}(\vec{q})$ in first order. In second
and higher order products of the rotation matrices occur. In the
$\mu$-th order contribution of type
\begin{eqnarray}
& &\frac{1}{V^\nu} \sum_{q_1...q_\nu} \sum_{\kappa_1...\kappa_\nu
\atop \kappa'_1...\kappa'_\nu} 
\tilde{v}_{\kappa_1, \kappa'_1}(\vec{q}_1) \cdot ... \cdot
\tilde{v}_{\kappa_\nu, \kappa'_\nu}(\vec{q}_\nu) \; \cdot \nonumber \\
& & \; \cdot \qquad e^{-i(\vec{q}_1 + ...
+ \vec{q}_\nu) (\vec{x}_j - \vec{x}_{j'})} \; \cdot \nonumber \\
& & \; \cdot \;\; D_{\kappa_1}(\Omega_j) \cdot ... \cdot
D_{\kappa_\nu}(\Omega_j) \; \cdot
D_{\kappa'_1}^\ast(\Omega_{j'}) 
\cdot ... \cdot D_{\kappa'_\nu}^\ast(\Omega_{j'})
\label{eq:app22}
\end{eqnarray}
occur with $\nu \le \mu$. For $\nu = 1$ it coincides with the first
order contribution. Using the product rule for the $D$'s, we get for
$\nu = 2$ a contribution to $S_{\kappa,\kappa'}^{(d)}(\vec{q})$ which
is proportional to:
\begin{eqnarray}
A_{\kappa \kappa'}^{(2)}(\vec{q}) &\equiv& \frac{1}{V} {\sum_{q_1 q_2}}'
\sum_{\kappa_1 \kappa'_1 \atop \kappa_2 \kappa'_2} {\cal C}(\kappa_1
\kappa_2 \kappa) {\cal C}(\kappa'_1 \kappa'_2 \kappa') 
\; \cdot \nonumber \\
& & \; \cdot \; \tilde{v}_{\kappa_1 \kappa'_1}(\vec{q}_1) \tilde{v}_{\kappa_2
\kappa'_2}(\vec{q}_2)
\label{eq:app23}
\end{eqnarray}
where $\sum'_{q_1 q_2}$ denotes summation over $\vec{q}_1$ and
$\vec{q}_2$ such that $\vec{q}_1 + \vec{q}_2 = \vec{q}$ and
\begin{equation}
{\cal C}(\kappa_1 \kappa_2 \kappa) = C(l_1 l_2 l;m_1 m_2 m) C(l_1
l_2 l;n_1 n_2 n)
\label{eq:app24}
\end{equation}
is a product of Clebsch Gordan coefficients. Then it follows from
(\ref{eq:app23}) with the properties of $C(l_1 l_2 l;n_1 n_2 n)$
\cite{G&G}:
\begin{eqnarray}
A_{\underline{\kappa} \kappa'}^{(2)}(\vec{q}) &\equiv& 
\frac{1}{V} {\sum_{q_1 q_2}}'
\sum_{\kappa_1 \kappa'_1 \atop \kappa_2 \kappa'_2} 
(-1)^{l_1+l_2+l} {\cal C}(\kappa_1
\kappa_2 \kappa) {\cal C}(\kappa'_1 \kappa'_2 \kappa') \; \cdot \nonumber \\ 
& & \; \cdot \; \tilde{v}_{\kappa_1 \kappa'_1}(\vec{q}_1) \tilde{v}_{\kappa_2
\kappa'_2}(\vec{q}_2)
\label{eq:app25}
\end{eqnarray}
where we have used the independence of $\tilde{v}_{\kappa_1
\kappa'_1}$ and of $\tilde{v}_{\kappa_2 \kappa'_2}$ on the sign of
$n_1$ and $n_2$. First of all we observe that all terms in the sum of
(\ref{eq:app25}) with $l_1+l_2+l$ even coincide with the corresponding
terms in (\ref{eq:app23}), independent of $\vec{q},m_1$ and $m_2$. Hence
for those terms the $n~\to~-n$ symmetry holds for
arbitrary $\vec{q},l_1,m_1,l_2$ and $m_2$, provided $l_1+l_2+l$ is
even. On the other hand choosing $m=0$ and $\vec{q} = \vec{q}_0 =
(0,0,q)$ ($q$--frame) and changing $m_1 \to -m_1, \; m_2 \to -m_2$ we
find 
\begin{eqnarray}
A_{\underline{\kappa} \kappa'}^{(2)}(\vec{q}) &=&
\frac{1}{V} {\sum_{q_1 q_2}}'
\sum_{\kappa_1 \kappa'_1 \atop \kappa_2 \kappa'_2}
{\cal C}(\kappa_1 \kappa_2 \kappa) {\cal C}(\kappa'_1 \kappa'_2
\kappa') \; \cdot \nonumber \\
& & \; \cdot \; \tilde{v}_{\overline{\kappa}_1 \kappa'_1}(\vec{q}_1)
\tilde{v}_{\overline{\kappa}_2 \kappa'_2}(\vec{q}_2)
\label{eq:app26}
\end{eqnarray}
where $\overline{\kappa} = (l,\underline{m},n)$. The $(\vec{q}_1 m_1,
\vec{q}_2 m_2)$--dependence of $\tilde{v}_{\overline{\kappa}_1
\kappa'_1}(\vec{q}_1) \tilde{v}_{\overline{\kappa}_2
\kappa'_2}(\vec{q}_2)$ is given by (cf. (\ref{eq:app14}) ):
\begin{equation}
Y_{l_1 \underline{m}_1}(\vec{e}_{q_1}) Y_{l_2
\underline{m}_2}(\vec{e}_{q_2}) \propto e^{-i(m_1 \phi_1 + m_2 \phi_2)}
= e^{i m_1 (\phi_2 - \phi_1)}
\label{eq:app27}
\end{equation}
where $0 = m = m_1 + m_2$ has been used. The azimuthal angles $\phi_1$
and $\phi_2$ of $\vec{e}_{q_1}$ and $\vec{e}_{q_2}$ fulfill $\phi_2 -
\phi_1 = \pi$ due to $\vec{q}_0 \vec{e}_{q_1} + \vec{q}_0
\vec{e}_{q_2} = 0$, i.e. the r.h.s. of (\ref{eq:app27}) equals
$(-1)^{m_1}$ which does not depend on the sign of $m_1$. Therefore we
find that $Y_{l_1 \underline{m}_1}(\vec{e}_{q_1}) Y_{l_2
\underline{m}_2}(\vec{e}_{q_2}) = Y_{l_1 m_1}(\vec{e}_{q_1}) Y_{l_2
m_2}(\vec{e}_{q_2})$ and accordingly that
\begin{equation}
\tilde{v}_{\overline{\kappa}_1 \kappa'_1}(\vec{q}_1)
\tilde{v}_{\overline{\kappa}_2 \kappa'_2}(\vec{q}_2) =
\tilde{v}_{\kappa_1 \kappa'_1}(\vec{q}_1) \tilde{v}_{\kappa_2
\kappa'_2}(\vec{q}_2).
\label{eq:app28}
\end{equation}
Making use of (\ref{eq:app28}), we get from (\ref{eq:app26}) that
\begin{equation}
A_{\kappa \kappa'}^{(2)} (\vec{q}_0) = A_{\underline{\kappa}
\kappa'}^{(2)} (\vec{q}_0)
\label{eq:app29}
\end{equation}
for all $\kappa'$ and $\kappa$ with $m=0$. We see that the $n \to -n$
(or $n' \to -n'$) symmetry of all the second order contributions only
holds in the $q$--frame and for $m = m' = 0$. 

Besides $A_{\kappa \kappa'}^{(2)} (\vec{q})$, in third order there
also exist contributions of type
\begin{eqnarray}
A_{\kappa \kappa'}^{(3)} (\vec{q}) &=& \frac{1}{V^2} 
{\sum_{q_1 q_2 q_3}}'
\sum_{\kappa_1 \kappa_2 \kappa_3 \atop \kappa'_1 \kappa'_2 \kappa'_3}
\; \cdot \nonumber \\
& & \; \cdot \;\; ( ... ) \tilde{v}_{\kappa_1 \kappa'_1}(\vec{q}_1)
\tilde{v}_{\kappa_2 \kappa'_2}(\vec{q}_2) \tilde{v}_{\kappa_3
\kappa'_3}(\vec{q}_3)
\label{eq:app30}
\end{eqnarray}
where $( ... )$ contains products of ${\cal C}$'s, and the sum over
$\vec{q}_1, \vec{q}_2$ and $\vec{q}_3$ is restricted to
$\vec{q}_1+\vec{q}_2+\vec{q}_3 = \vec{q}$. For $\vec{q} = \vec{q}_0$
and $m = 0$ the same procedure as for $\nu = 2$ can be used in order
to prove that the $n \to -n$ symmetry will be true if the three--fold
product of the $\tilde{v}$'s in (\ref{eq:app29}) is invariant under
$m_i \to -m_i, \; i = 1,2,3$. This requires that (in analogy to $\nu =
2$):
\begin{equation}
e^{-i(m_1 \phi_1 + m_2 \phi_2 + m_3 \phi_3)} = e^{i(m_1 \phi_1 + m_2
\phi_2 + m_3 \phi_3)}.
\label{eq:app31}
\end{equation}
Although $0 = m = m_1+m_2+m_3$, the azimuthal angles $\phi_i$ of
$\vec{e}_{q_i}$ (in contrast to $\nu=2$) are not restricted because
$\sum_{i=1}^{3} \vec{q}_0 \cdot \vec{q}_i = 0$ does not fix $\phi_3$ if
$\phi_1$ and $\phi_2$ are given. Therefore the generic case for
$\vec{q}_i$ obeying $\sum_{i=1}^{3} \vec{q}_0 \cdot \vec{q}_i = 0$
will not satisfy (\ref{eq:app30}). But similar to $\nu = 2$, there is
a subset of terms in the sum of (\ref{eq:app29}) which will be
invariant under $n \to -n$ (or $n' \to -n'$) for {\em all}
$\vec{q},l,m,l'$ and $m'$.

The results show that the invariance of $S_{lmn,l'm'n'}^{(d)}(\vec{q})$
for arbitrary $(\vec{q},l,m,l',m')$ and for
$(\vec{q}=\vec{q}_0,l,l',m=m'=0)$ only holds up to, respectively, the
first and second order in $1/k_BT$.

\end{appendix}


\begin{references}
\bibitem{note} The ionic glassformer CKN could be considered 
as an exception, although it contains
molecular units like $NO_3$--ions.
\bibitem{predictions}
The validity of the
factorization in $i)$ is guaranteed when the largest eigenvalue $E_0$
of the stability matrix (see e.g. \protect\cite{leshouches} for simple 
liquids) at the critical temperature related
to the nonlinear equation for the nonergodicity parameter is {\em not}
degenerate. 
In case that the vertices of the memory kernel which
enters the MCT equations are positive this is automatically fulfilled.
Since the vertices occurring in MMCT are neither real nor positive this
nondegeneracy must not hold necessarily. Nevertheless, we expect that
$E_0$ in the generic case is nondegenerate. An investigation of $E_0$
for a system of diatomic molecules confirms this statement 
\protect\cite{winkler}.
The second scaling law is easily proved by showing that the Laplace
transformed MMCT equations for $\hat{S}_{lmn,l'm'n'}(\vec{q},z)$ are
invariant under the rescaling $z \to yz$ for any real
number $y$,
provided the microscopic frequencies are neglected.
\bibitem{rotation} Properties of the 
molecular correlators  under rotations for the case of
linear molecules have been discussed in 
\protect\cite{dumbbell,schilling} and for arbitrary molecules in 
\protect\cite{theisdipl}.
\bibitem{osc} The minimum seen in all correlators around
0.5 ps is a finite-size effect connected to
the low frequency cut-off the density of states imposed by the
boundary conditions and/or by the undamped propagation of the
sound waves through the boundary conditions. 
Enhanced oscillations in the  $ps$-region have been
observed in several molecular dynamics studies of supercooled liquids, 
especially close to the ideal glass transitions and
finite-size-effect studies have shown that
these oscillations are spurious. 
In network forming liquids there is evidence 
that the finite-size-effects may be superimposed to a 
much weaker oscillation
associated to the time domain manifestation of the boson-peak.
\end{references}
\end{document}